\documentclass[journal=jpclcd,manuscript=article,articletitle=true,layout=twocolumn]{achemso}

\usepackage{amsmath}
\usepackage{amssymb}
\usepackage{latexsym}
\usepackage{graphicx}
\usepackage{epsfig,epsf,rotating}
\usepackage{indentfirst}
\usepackage{epstopdf}
\usepackage{xcolor}
\usepackage{multirow}
\usepackage{tabularx}
\usepackage{ctable}
\usepackage{textcomp}
\usepackage[version=4]{mhchem}
\usepackage{amsmath}

\usepackage{upgreek}
\newcolumntype{C}{>{\centering\arraybackslash}X}
\usepackage{nicefrac}

\title{Polarizability Plays a Decisive Role in Modulating Association Between Molecular Cations and Anions\footnote{
Notice:  This manuscript has been authored by UT-Battelle, LLC, under contract DE-AC05-00OR22725 with the US Department of Energy (DOE). The US government retains and the publisher, by accepting the article for publication, acknowledges that the US government retains a nonexclusive, paid-up, irrevocable, worldwide license to publish or reproduce the published form of this manuscript, or allow others to do so, for US government purposes. DOE will provide public access to these results of federally sponsored research in accordance with the DOE Public Access Plan (http://energy.gov/downloads/doe-public-access-plan).}}

\author{Chase~E.~Herman}
\affiliation{Department of Chemical and Biomolecular Engineering, 150 Academy St., University of Delaware, Newark, DE 19716, USA}

\author{Arjun~Valiya~Parambathu}
\affiliation{Department of Chemical and Biomolecular Engineering, 150 Academy St., University of Delaware, Newark, DE 19716, USA}

\author{Dilip~N.~Asthagiri}
\affiliation{Oak Ridge National Laboratory, 1 Bethel Valley Rd., Oak Ridge, TN 37830, USA}
\email{asthagiridn@ornl.gov}

\author{Abraham~M.~Lenhoff} %  \corref{cor1}
\affiliation{Department of Chemical and Biomolecular Engineering, 150 Academy St., University of Delaware, Newark, DE 19716, USA}
\email{lenhoff@udel.edu}

\begin{document}
    \begin{abstract}
Electrostatic interactions involving proteins depend not just on the ionic charges involved but also on their chemical identities. Here we examine the origins of incompletely understood differences in the strength of association of different pairs of monovalent molecular ions that are relevant to protein-protein and protein-ligand interactions. Cationic analogues of the basic amino acid side chains are simulated along with oxyanionic analogues of cation-exchange (CEX) ligands and acidic amino acids. Experimentally observed association trends with respect to the cations, but not anions, are captured by a non-polarizable model. A polarizable model proves decisive in capturing experimentally-suggested trends with respect to both cations and anions. Crucially, relative to a non-polarizable model, polarizability changes the free energy surface for ion-pair association, altering configurational sampling itself. An effective continuum correction to account for electronic polarizability can also capture the experimentally-suggested trends, but at the expense of fidelity to the underlying free energy surface. 

    \begin{tocentry} 
        \includegraphics[height=5cm]{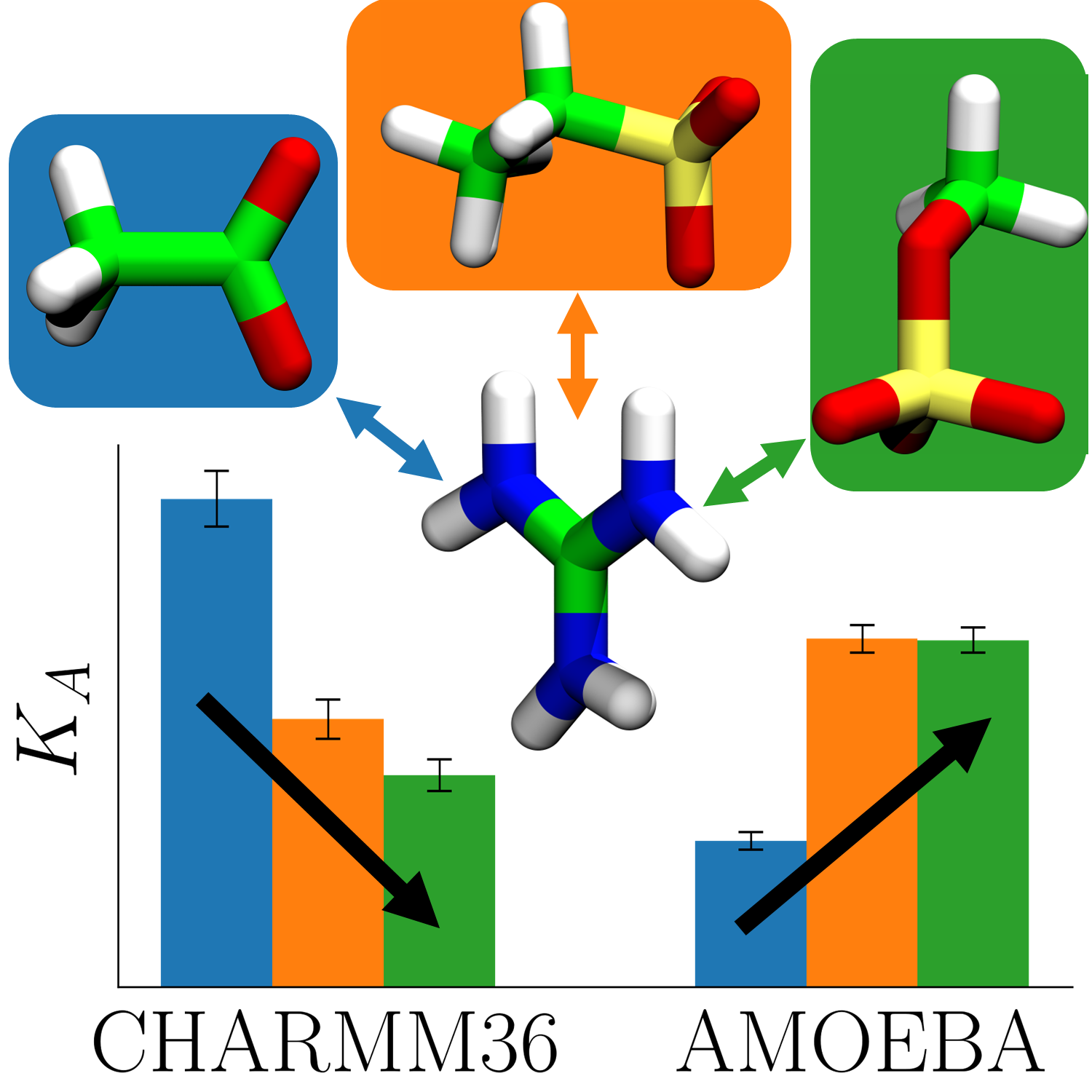} 
    \end{tocentry}
    
    \end{abstract}

    \maketitle

    Interactions among charged groups in aqueous solution are widely implicated in protein structural stabilization, biomolecular recognition and surface adsorption.\cite{Haggerty1991, Roberts2014, Luo1999} Among the specific phenomena of interest to this work is the longstanding puzzle that protein adsorption in cation-exchange (CEX) chromatography is known to be stronger on resins that are decorated with sulfate or sulfopropyl ligands than on resins with carboxymethyl ligands, despite the fact that the ligands bear the same formal charge and may be present on equivalent base matrices at comparable immobilization densities.\cite{DePhillips2001, Asthagiri2000} A related puzzle is the stronger binding to a heparin affinity resin of arginine than lysine 7-mers.\cite{Fromm1995} Uncovering the physics that underlie these observations on systems of broad scientific and technological interest still remains largely an open problem. 

    Given the ionic nature of the constituents noted above, it is a natural first step to model the physics relying solely on electrostatics whilst ignoring the molecular nature of the solvent. Continuum solvent approaches have been enormously influential in guiding our thinking but they treat hydration phenomena only approximately and are unable to resolve puzzles such as those noted above\cite{Warshel2006, Ren2012, Koehl2006, Gray2018, Sun2016, Asthagiri2021b, Gunsteren2006}. It is necessary to account better the underlying physics. 

    \begin{figure}[ht]
        \begin{center}
            \includegraphics[width=\columnwidth]{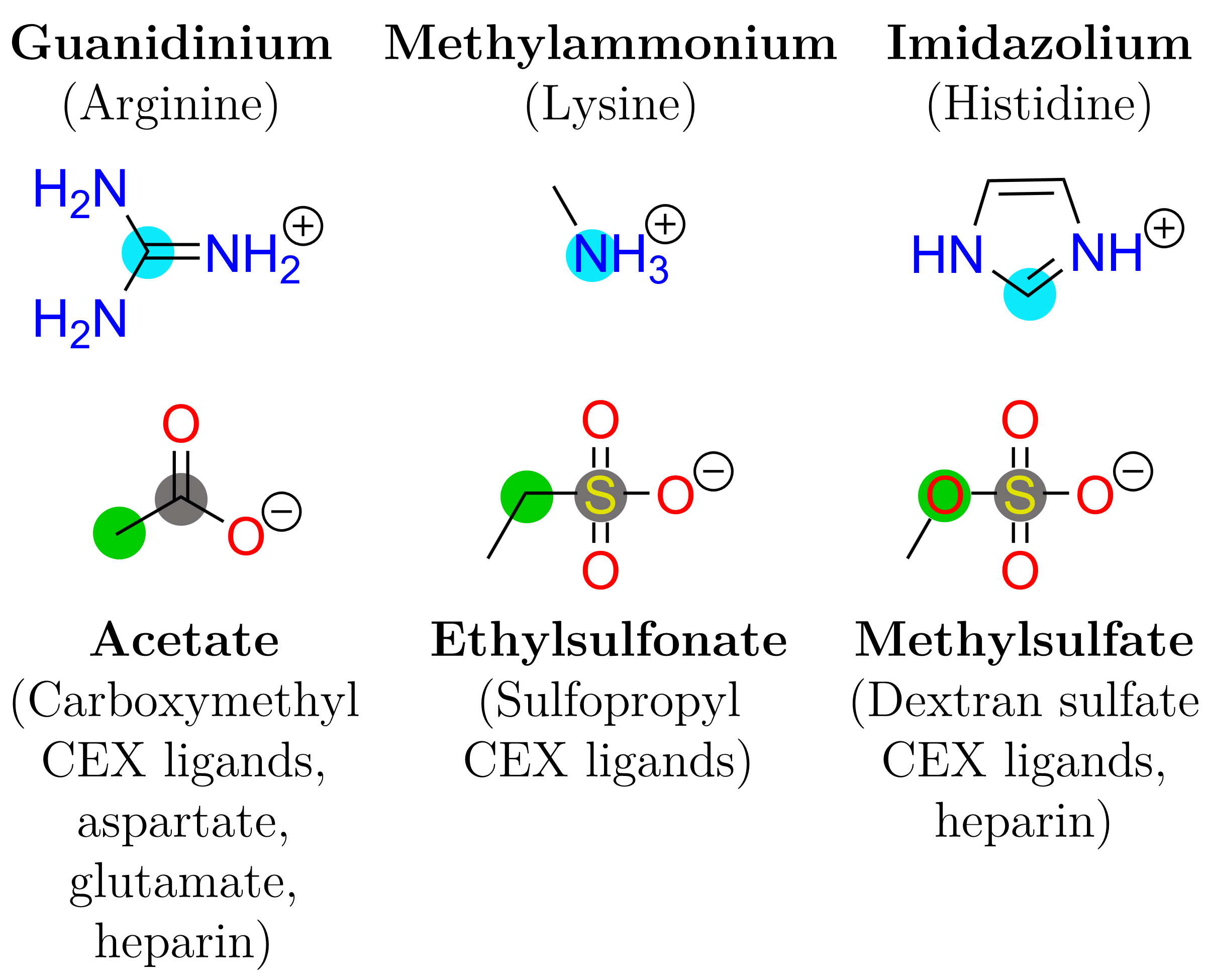}
            \caption{Cationic analogues (top) of the basic amino acid side chains and oxyanionic analogues (bottom) that were simulated in this work and  are relevant to cation-exchange (CEX) chromatography, protein-protein interactions and protein-heparin interactions. Ion-pair separation distances $r$ were measured between the atoms that are highlighted in blue and grey circles and the coordinate $\theta$ was defined as the angle between the atoms that are highlighted in blue, grey and green circles.} 
            \label{fig:schematic}
        \end{center}
    \end{figure}
    
    All-atom molecular dynamics (MD) can in principle capture the balance between direct solute-solute and indirect solvent-mediated intermolecular interactions.  However, classical non-polarizable force fields (FFs) are generally known to overestimate the strength of ion-pair interactions.\cite{Debiec2014, Debiec2016, Mason2019a} This may be attributable to the partial inclusion of polarization implicitly in the parameterization of water models but not in ion models.\cite{Dijon2020} To fix this disparity in the treatment of the solvent versus the ion, it has been suggested that ion charges be scaled by a constant factor of $1/\sqrt{\varepsilon_{el}} \sim 0.75$, where $\varepsilon_{el} \sim 1.78$ represents the high-frequency dielectric constant of water.\cite{Leontyev2011, Dijon2020}. Results using this so-called  electronic continuum correction (ECC) appear to better capture the strength of ion-pairing in solution. 
    However, relative to the parent FF, the ECC complicates the estimation of solute hydration free energies \cite{Dijon2020} and may err in describing intermolecular interactions at large separations. 

      \begin{figure*}[ht!]
        \begin{center}
            \includegraphics[width=\textwidth]{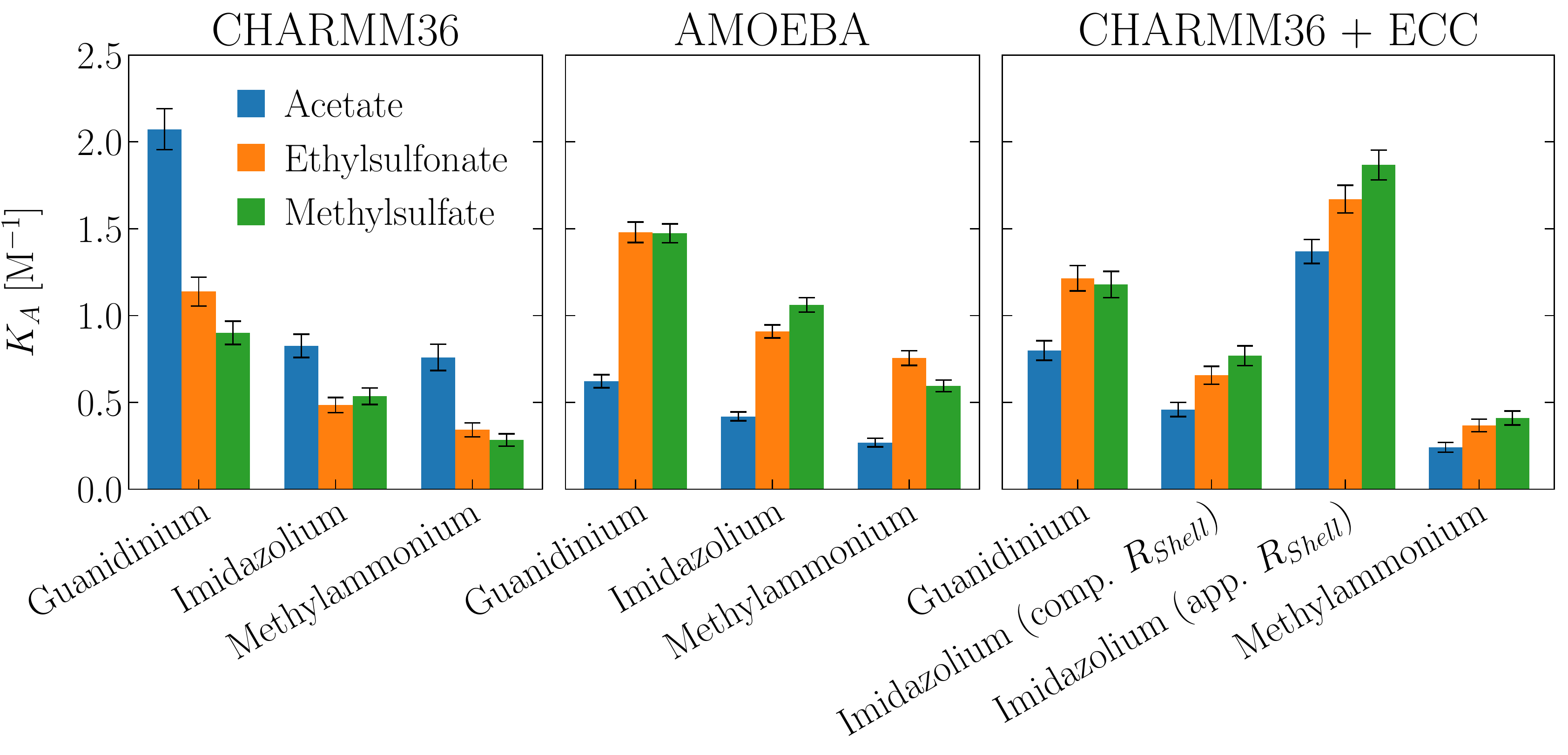}
            \caption{Association constants obtained from simulated interaction of cations (abscissa) with anions (bars) at a concentration of $\sim$ 0.5 M using different force fields (panels). ECC refers to the electronic continuum correction and error bars represent $2\times$ the standard error of the mean. For CHARMM36 + ECC imidazolium systems, ``comp." and ``app." refer to comparable and apparent $R_{Shell}$ values that correspond to contact pair and solvent-separated PMF minima (Fig.~S1). Data are given in Table~S1.} 
            \label{fig:Ka_bar_plot}
        \end{center}
    \end{figure*}	 
  
    Here we assess the ability of a classical FF (CHARMM36\cite{Huang2013}) with and without the ECC as well as a semiclassical model (AMOEBA\cite{Shi2013}) to explain the two puzzles noted above. (We describe the AMOEBA model as semiclassical because of its rather detailed description of electrostatics and its inclusion of polarizability.) Our previous work treated the solute at a quantum chemical level and
    the solvent as a continuum;\cite{Asthagiri2000} that effort captured some of the effects qualitatively, albeit with much uncertainty given the study's limited configurational exploration. Following that work, here we model analogues of oxyanionic CEX ligands (i.e., acetate, methylsulfate and ethylsulfonate for carboxymethyl, dextran sulfate and sulfopropyl resins, respectively) interacting with comparable analogues of the basic amino acid side chains (i.e., guanidinium, methylammonium and imidazolium for arginine, lysine and histidine residues, respectively) (Fig. \ref{fig:schematic}); acetate is of course also a model for the acidic amino acid side chains. Focusing on ligands alone allows us to study physically critical interactions more thoroughly. The insights thus derived also have broad relevance to modeling protein interactions as well as ionic liquids.

   \textbf{Association constants:} Unbiased simulations were performed for each anion-cation pair at a concentration of $\sim$ 0.5 M to compute potential of mean force (PMF) profiles (Supplementary Figure~S1); association constants $K_A$ were calculated as\cite{Zhu2012, Shi2017}
    \begin{eqnarray}
    K_{A} = \frac{1}{K_{D}} = 4 \pi \int_{0}^{R_{Shell}} r^2 e^{-w(r)/k_B T} \; dr
    \label{eqn:Ka}
    \end{eqnarray}
    where $r$ and $w$ refer to the separation distance and the PMF, respectively, and the ions are considered to be associated for $r < R_{shell}$.
    The location of the first maximum in the PMF was chosen as $R_{shell}$,\cite{Shi2017} which separates the contact ion-pair from the solvent-separated ion pair and occurs at $r \sim 6$~{\AA} (Table~S1). 

Figure~\ref{fig:Ka_bar_plot} shows that both CHARMM36 and AMOEBA capture the expected $K_A$ rank order in cations (i.e., guanidinium $>$ imidazolium $>$ methylammonium) based on retention data of individual amino acids on sulfonate resins;\cite{Wang1989, Moore1958} the order is also consistent with the Lys and Arg oligomer retention in heparin affinity chromatography.\cite{Fromm1995} Although confounded by protein structural details, this order is furthermore consistent with the stronger retention of lysozyme (which is Arg-rich) than cytochrome \emph{c} (which is Lys-rich) on several CEX resins, which continuum electrostatics models may not be able to capture.\cite{DePhillips2001, Yao2005} CHARMM36 with the ECC generally captures the cation order, although the ECC tends to understructure the PMF profiles relative to AMOEBA (Fig.~S1). This is especially the case for imidazolium systems, making the definition of $R_{Shell}$ somewhat unclear. For this reason, comparable and apparent $R_{Shell}$ values are used in Figure~\ref{fig:Ka_bar_plot} that correspond approximately to contact and solvent-separated ion pairs.

The more revealing comparisons are among the anions: CHARMM36 \emph{incorrectly} predicts stronger cation interactions with acetate than with the sulfur-containing ligands. AMOEBA and the ECC both correct this trend to be consistent with the experimentally observed rank order of CEX resin retentivities. Two competing features stand out. From the perspective of the ions, relative to AMOEBA, CHARMM36 \emph{over}-stabilizes the interactions with acetate and \emph{under}-stabilizes the interactions with the sulfur-containing anions  (Fig.~S1). From the perspective of the solvent, relative to AMOEBA, CHARMM36 predicts \emph{weaker} acetate-water interactions but \emph{stronger} water interactions with the sulfur-containing anions (Fig.~S3). 
The ECC brings the classical model $K_A$ predictions closer to those of AMOEBA but sometimes at the expense of fidelity to the underlying PMF for ion-ion and ion-water association (Figs.~S1 and~S3).  However, this observation must be tempered by the recognition that it is likely necessary to update the water models themselves for use in the ECC-ecosystem \cite{Dijon2020}. 

    There is a paucity of data to support more rigorous quantitative comparisons between these results and experiment, but guanidinium acetate association has been investigated in two independent potentiometric studies.\cite{Tanford1954, Haake1977} Association constants were inferred in those studies from an induced shift in the ionization constant of acetic acid when guanidinium was substituted for another cation that was assumed not to complex with acetate. $K_A$ was independently estimated to be $\sim$~0.5~M$^{-1}$ and 0.37~M$^{-1}$,\cite{Tanford1954, Haake1977} but the ionic strength (IS) was uncontrolled in the first study,\cite{Tanford1954} which obfuscates the result because $K_A$ is expected to decrease with IS due to electrostatic screening. The IS was fixed at 1.02~M in the second study\cite{Haake1977} and the experimental value of 0.37~M$^{-1}$ is of comparable magnitude but appropriately lower than the AMOEBA prediction of 0.62~M$^{-1}$, which was obtained for guanidinium acetate at an IS of $\sim$~0.5~M. A comparable measurement of the interaction between butylammonium and acetate (0.31~M$^{-1}$ at 1.02~M IS)\cite{Haake1977} is also similar to the AMOEBA prediction for the methylammonium acetate system (0.27~M$^{-1}$ at $\sim$~0.5~M IS). The data from acetate simulations are similar to previous computational results as well,\cite{Mason2019a, Debiec2014, Debiec2016} including a difference that was observed in classical simulations of guanidinium acetate using a different FF with and without the ECC (Fig.~S4).\cite{Mason2019a}
    
    \textbf{Cross-FF analysis:} To assess whether the dissimilarities between the CHARMM36 and AMOEBA results were due to polarizability or simply a difference in the treatment of electrostatics involving permanent charges, a cross-FF analysis was performed in which the \emph{in vacuo} interaction energies between all ion pairs in AMOEBA simulation trajectories were analyzed retrospectively using both the CHARMM36 and AMOEBA FFs. This permitted an applied comparison of FF parameterization without any differences in configurational sampling. For each FF used in the analysis, the data were ensemble-averaged as a function of the separation distance $r$; Figures~S5-S7 show the decomposition of the \emph{in vacuo} interaction energy into permanent electrostatic ($U_{Elect}$), van der Waals ($U_{VdW}$) and polarization ($U_{Polar}$) contributions. Figure~S8 shows the magnitude of the difference between the CHARMM36 and AMOEBA profiles. 

    \begin{figure*}[ht]
    \begin{center}
        \includegraphics[width=\textwidth]{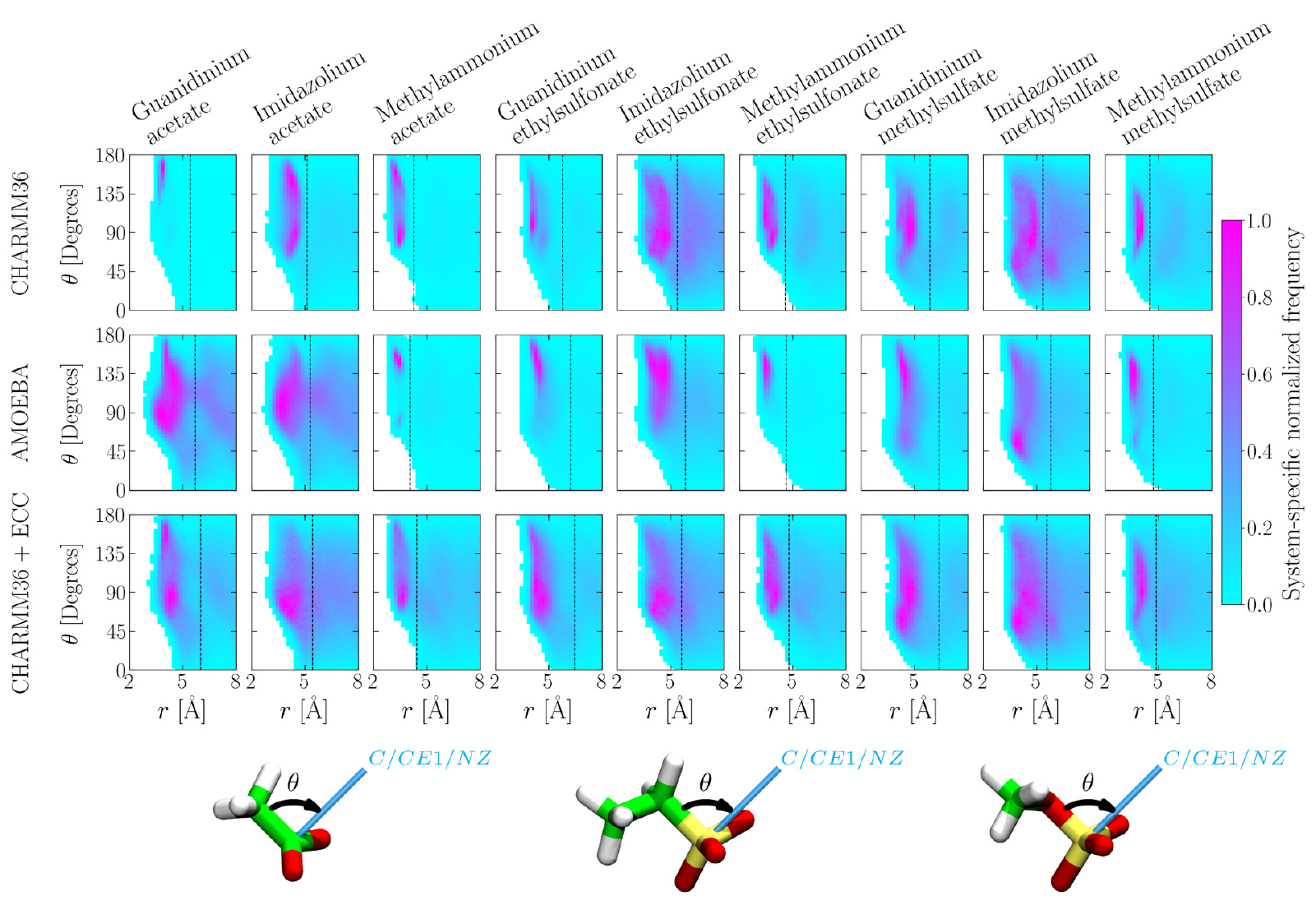}
        \caption{Heat maps of the sampling frequency of configurations defined by the coordinates $(r, \theta)$ for each cation-anion pair (columns) simulated using each force field (rows). Dashed lines represent $R_{Shell}$ and the normalization of sampling frequency is specific to each system. (PMF profiles may be recovered as a projection of these data prior to normalization.)}
        \label{fig:heatmaps_theta_r}
    \end{center}
    \end{figure*}
    
The permanent electrostatic contributions are nearly identical for the two FFs, with the largest difference observed in methylsulfate systems, where AMOEBA predicts slightly stronger attraction than CHARMM36 (Fig.~S5). This similarity of electrostatics is intriguing because AMOEBA uses a sophisticated distributed multipole model whereas CHARMM36 uses only point charges at atom centers.\cite{Huang2013, Shi2013}  AMOEBA predicts generally weaker
van der Waals interactions than CHARMM36 but the magnitude of the difference is relatively small (Fig.~S6). Polarization contributions represent the largest difference between the two FFs (Fig.~S7). Thus while AMOEBA's more sophisticated multipole treatment of permanent electrostatics does contribute to sampling differences,\cite{Shi2013} polarizability is the primary factor that is responsible for differences in model performance. Polarization contributions are always observed to stabilize counterion pairs \emph{in vacuo}. 

    \textbf{PMF decomposition:} The PMF may be decomposed as $w = U_{Total} + W_{Solv}$, 
    where  $U_{Total} = U_{Elect} + U_{VdW} + U_{Polar}$ is the \emph{in vacuo} interaction energy and $W_{Solv}$ is the solvent-mediated
    contribution. The contributions to $U_{Total}$ were estimated using the trajectory-generating FF and $W_{Solv}$ was found from the PMF by difference; the results are shown in Figures S9-S12. Near the PMF minima, the magnitudes of both $U_{Total}$ and $W_{Solv}$ are on the order of 90 kcal/mol (Figs.~S9 and S10). Thus, it is the fine balance between two large competing 
    contributions that dictates the PMF,\cite{Shi2017} which has a well depth that is typically two orders of magnitude smaller.

    Given the similarities that were observed in the cross-FF analysis, differences between the CHARMM36 and AMOEBA profiles in Figures~S9-S12 may be primarily attributed to the sampling differences that polarizability promotes. Permanent electrostatic interactions, which comprise the principal contribution to $U_{Total}$, are similar for CHARMM36 and AMOEBA in most systems (Fig.~S11). However, noticeable discrepancies are apparent in the guanidinium acetate and imidazolium acetate systems, for which CHARMM36 substantially overestimates the magnitude of $U_{Elect}$. The favorable AMOEBA polarization contributions (Fig.~S7) decrease this discrepancy in the $U_{Total}$ profiles for imidazolium acetate (Fig.~S9) but a substantial difference remains for guanidinium acetate. In general, polarization contributions for the other systems and differences in permanent electrostatics lead to more attractive $U_{Total}$ profiles for AMOEBA than CHARMM36 (Fig.~S9). The inferred solvent contribution $W_{Solv}$
 necessarily follows the opposite trend (Fig.~S10).

    \textbf{Configurational sampling:}  Differences in configurational sampling were examined more directly for each system by identifying the 3D spaces of the most frequently observed cation positions relative to anions (Fig.~S14) and anion positions relative to cations (Fig.~S15). For example, for guanidinium interacting with acetate or ethylsulfonate (Fig.~S14), the cation samples a small patch of space around the carboxylate group within the CHARMM36 description but this is relaxed in the AMOEBA description.  This trend is approximately reversed for 
    ethylsulfonate. With the ECC correction, no such distinction is evident, suggesting a more promiscuous sampling and hence weaker association.  Similar distinctions can be noted from either the anion's (Fig.~S14) or the cation's perspective (Fig.~S15). To quantify better the sampling differences suggested by the 3D plots, we project the data on an $(r,\theta)$ map (Fig.~\ref{fig:heatmaps_theta_r}; $\theta$ is defined in Fig.~\ref{fig:schematic}). For $\theta$ between 90\textdegree \ and 180\textdegree, the carbonyl or sulfonyl oxygen is directed towards the anion.  
    
Summarizing the observations, in general, for a given cation, CHARMM36 predicts that configurational preferences become less well-defined as the number of oxygen atoms in the anion increases, i.e., as the charge density decreases. This is as expected on the basis of the dominance of electrostatic interactions.  With some exceptions, the opposite holds for AMOEBA, suggesting a richer interplay of electrostatic interactions arising from charge-charge interactions and polarization effects. The effect of the ECC is inconsistent. 
    
 \textbf{Hydration free energies:} Emergent behavior in ion-pair association is a consequence of the balance between direct interactions 
    and solvent-mediated effects that generally tend to suppress ion-pair formation. The latter arises because there is an 
    energetic penalty for partially dehydrating the interacting ions and reorganizing the nearby solvent structure. Hydration free energies ($\mu^{\rm{ex}}$) may therefore provide a useful complement to inform the understanding of ion pairing phenomena. To this end we estimated $\mu^{\rm{ex}}$ for individual ions using the molecular quasichemical organization of the potential distribution theorem,\cite{Beck2006, Asthagiri2021b, Weber2011, Weber2012} which allows $\mu^{\rm{ex}}$ to be partitioned into physically meaningful contributions as
    \begin{eqnarray}
    \beta \mu^{\rm{ex}} = \underbrace{- \ln p_0[\phi]}_\text{Packing} + \underbrace{\beta\mu^{\rm{ex}}_{\rm{LR}}[P(\varepsilon \; | \; \phi)]}_\text{Long-range} + \underbrace{\ln x_0[\phi]}_\text{Chemistry}
    \label{eq:mqct}
    \end{eqnarray}
    where the first term represents the free energy required to open a cavity in the solvent to accommodate the ionic solute. Contributions from long-range and short-range solute-solvent interactions are given by the second and third terms, respectively. Each term is a functional of a repulsive potential $\phi$ that is used to condition the solvent up to a maximum range of $\lambda_G$ around the solute, but the sum of the three terms is independent of $\phi$. 
    We use $\lambda_G = 5$~{\AA}, for which the cavity corresponds roughly to the first hydration shell around the solute. 
 
    \begin{figure*}[ht]
        \begin{center}
            \includegraphics[width=\textwidth]{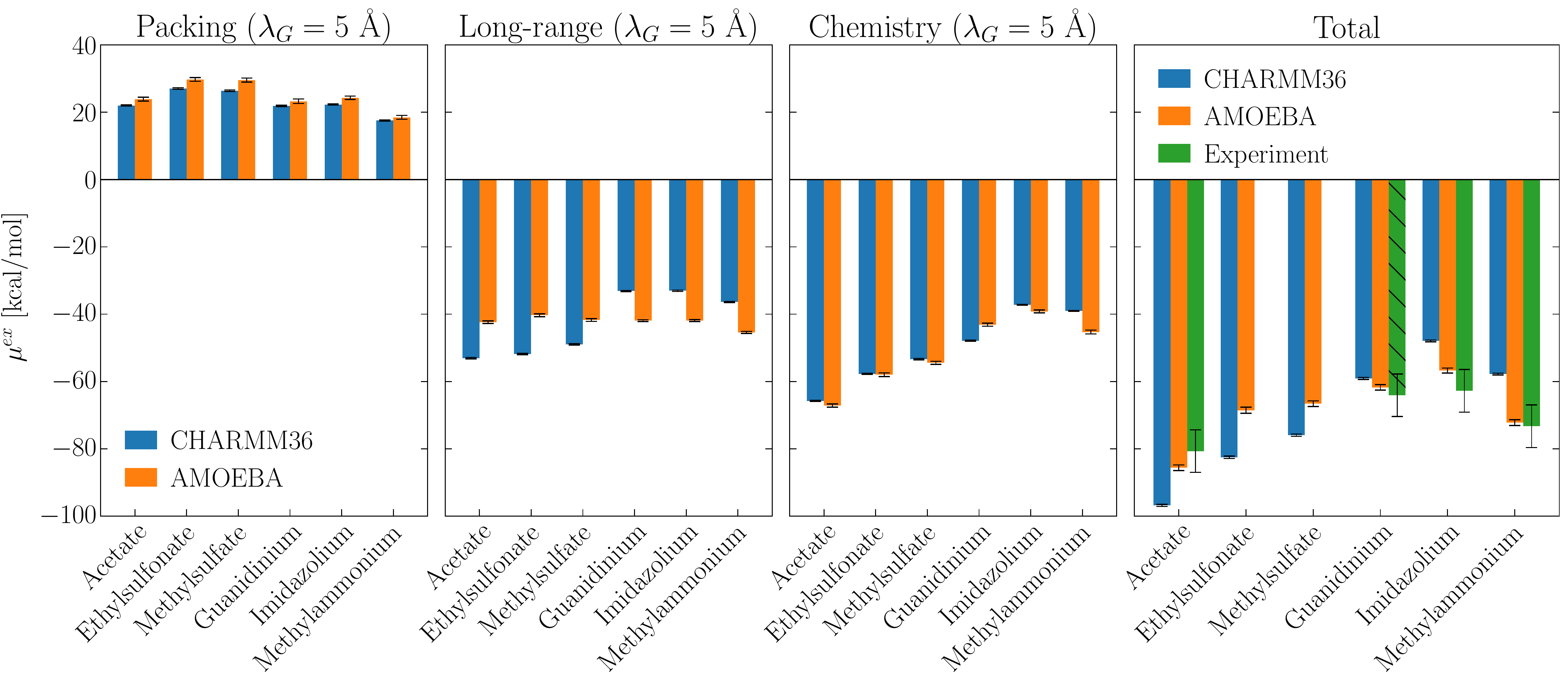}
            \caption{Partitioning of ion hydration free energies according to molecular quasichemical theory. $\lambda_G = 5$~{\AA} approximately defines the first hydration shell around the solute. Error bars on simulation results represent $2\times$ the standard error of the mean. Where available, experimental data were used to estimate $\mu^{\rm{ex}}$ using a thermodynamic cycle based on a proton hydration free energy of $-260.9 \pm 5.8$ kcal/mol, which represents the mean and standard deviation of 72 independent estimates.\cite{Fossat2021}. Data are given in Tables S3 and S4. Hatched lines indicate that the experimental estimate for guanidinium uses an approximation for the hydration free energy of guanidine (cf. Table S4).} 
            \label{fig:mqct}
        \end{center}
    \end{figure*}
 
 Experimental estimates of $\mu^{\rm{ex}}$ may be obtained using a thermodynamic cycle based on proton dissociation, which requires reference to the free energy of hydrating a proton ($\mu^{\rm{ex}}_{H^+}$).\cite{Fossat2021, Lim1991, Pearson1986} However, the value of $\mu^{\rm{ex}}_{H^+}$ is the subject of much uncertainty because extrathermodynamic assumptions must be employed to deconvolute experimentally accessible quantities into anion and cation contributions.\cite{Grossfield2003, Zhang2017, Fossat2021} We have taken $\mu^{\rm{ex}}_{H^+}$ and its uncertainty from a recent report that summarizes 72 independent estimates of the value.\cite{Fossat2021} Literature data exist to make experimental comparisons for the computed $\mu^{\rm{ex}}$ of acetate and the cations,\cite{Cramer1991, Taft1987, Cumming1977, Fujio1981, Settimo2014, Zhang2017, Wolfenden1981, Reif2012, Hunter1998, Rizzo2006, In2005} which are shown alongside simulation results in the rightmost panel of Figure~\ref{fig:mqct} and are detailed in Table~S4. These are comparable to other literature reports that are listed in Table~S5.\cite{Pearson1986, Cramer1991, Gilson1988, Marcus2013, Gokcen2014} Within the appreciable uncertainty, the AMOEBA results agree with experiment. Data for acetate and the cations are also similar to simulation results based on thermodynamic integration.\cite{Lin2018, Fossat2021, Zhang2017}   
 
 We next consider the individual $\mu^{\rm{ex}}$ contributions (Eq.~\ref{eq:mqct}). Uniformly, the packing contribution, i.e., the primitive hydrophobic contribution, to hydration is somewhat stronger (more positive) in AMOEBA than in CHARMM36. With the exception of guanidinium, the chemistry contribution is also stronger (more negative) in AMOEBA than in CHARMM36, which reflects greater ion-water attraction locally. Thus while packing will tend to favor ion-pair complexation in AMOEBA (over CHARMM36), the local attractive contributions will tend to suppress ion-pair formation in AMOEBA (over CHARMM36). 
  
  The long-range contribution to hydration proves surprising. This is more favorable for anions than cations in CHARMM36, which is expected based on the positive potential that exists in the center of a cavity due to the preferential orientation of water protons towards the cavity center \cite{Hummer1996,Ashbaugh2000}. However, the trend of long-range interactions found in AMOEBA is nearly reversed. 
  It is well-known that the sign and magnitude of the electric potential that the solvent imposes on the solute charges is sensitive to both the structure of the solvent at the interface and the description of the charge over the solvent molecules \cite{wilsonpratt:1988,doylebeck:2019}; for example, 
\emph{ab initio} simulations show that the dipole moment of solvent next to a large anion is itself reduced \cite{Guardia2009}. To probe this, in exploratory calculations for a spherical ion in water, we retrospectively included polarizability and multipole electrostatics in analyzing configurations sampled with a non-polarizable model. The shift is similar to the trend seen in Fig.~\ref{fig:mqct} (Long-range). A more thorough investigation is necessarily left for future studies.  
 
\begin{figure*}[ht]
        \begin{center}
            \includegraphics[width=\textwidth]{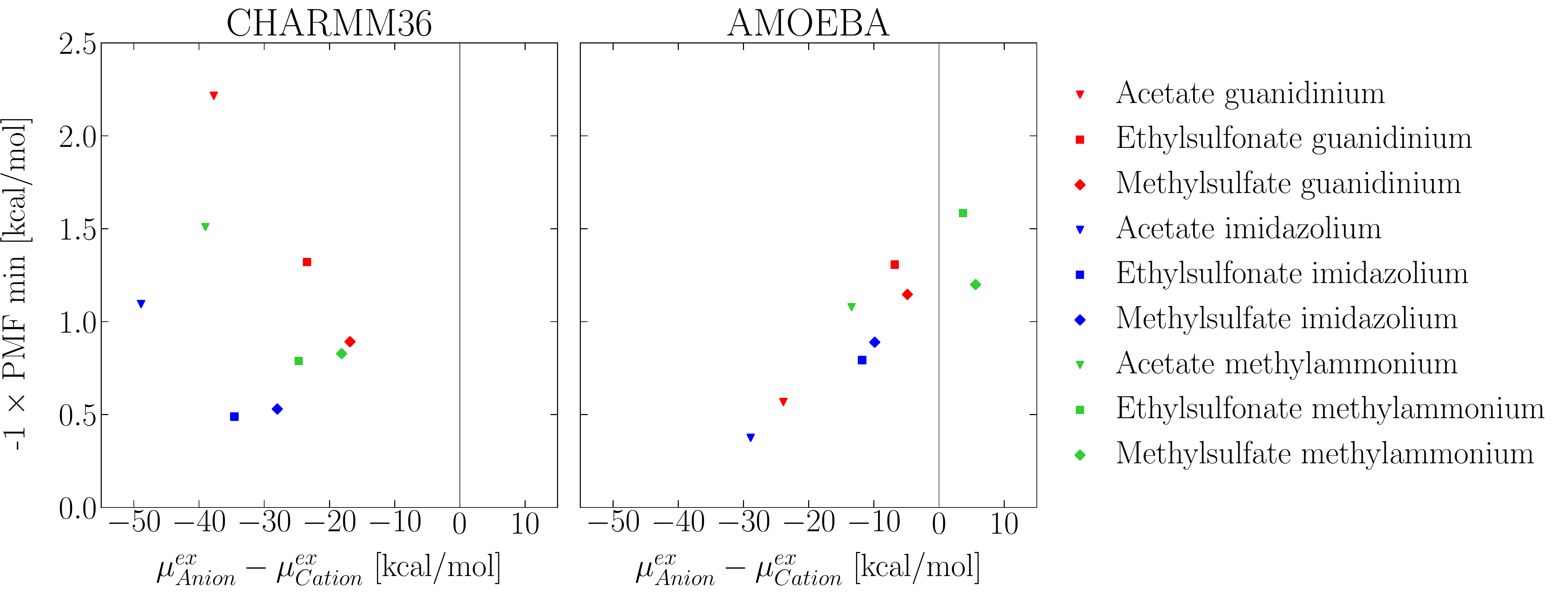}
            \caption{Comparison of potential of mean force well depths ($-1 \times \text{PMF min}$) with the difference in hydration free energies between the corresponding anion and cation as computed using CHARMM36 and AMOEBA. Marker sizes are close to $2\times$ the standard error of the mean (error not shown for marker readability).}
            \label{fig:volcano}
        \end{center}
    \end{figure*}

\textbf{Matching water affinities:}  To see if we can use the hydration free energies of individual ions to infer ion association, we test the applicability of the empirical law of matching water affinities\cite{Collins1997, Collins2019}. This rule suggests that ions with hydration free energies that are closely matched will tend to associate more readily. Figure~\ref{fig:volcano} shows that when using hydration free energies obtained from AMOEBA (which are closer to experimental estimates), there is a clear correlation between the difference in hydration free energy and the PMF well depth. The greater the difference in hydration free energy, the lower is the ion pair association affinity, which is reminiscent of experimentally observed trends in salt dissolution free energies.\cite{Collins1997} With CHARMM36 there is considerable scatter and the law of matching water affinities does not seem to hold. Although CHARMM36 predicts correctly that acetate is the most energetically expensive anion to dehydrate, it predicts incorrectly that cation interactions will be strongest with acetate. Overall, the results suggest that the empirical law of matching water affinities could be used in inferring ion-pairing, and hence potential protein adsorption on CEX matrices, provided the hydration of the individual ions is itself well captured (Fig.~\ref{fig:mqct}).

In conclusion, the present study suggests that the physics of polarizability is critical in determining why basic amino acid side chains bind more strongly to sulfate and sulfonate moieties than carboxylate groups, despite the fact that  carboxylate has a higher (negative) surface charge density and ought to interact more strongly with cations. The carboxylate moiety is more energetically expensive to dehydrate but predicting this is only half of the puzzle. Subtle polarization effects are required to capture experimental trends because it is the fine difference between two large competing potentials (i.e., electrostatic attraction and solvent opposition) that underlies ion complexation in solution. Although the ECC can sometimes improve this balance in classical FFs, it may also promote spuriously promiscuous configurational sampling.  Polarizability leads to qualitatively distinct configurational preferences that are expected to be broadly relevant to protein electrostatic interactions.

    \section{Supporting Information} 
        Supporting information includes methods and supplementary results for (1) unbiased simulations, (2) \emph{in vacuo} energy analyses (including cross-FF and PMF decomposition analyses), (3) configurational sampling analyses (including 3D maps, coordinate definitions and supplementary heat maps), (4) molecular quasichemical theory simulations and (5) experimental estimates of $\mu^{\rm{ex}}$ based on a thermodynamic cycle (PDF)
    
    \section{Acknowledgements}
        We thank Tom Beck (ORNL) for helpful discussions. 
        This research was supported in part through the use of DARWIN computing system: DARWIN – A Resource for Computational and Data-intensive Research at the University of Delaware and in the Delaware Region, which is supported by NSF under Grant Number: 1919839, Rudolf Eigenmann, Benjamin E. Bagozzi, Arthi Jayaraman, William Totten, and Cathy H. Wu, University of Delaware, 2021, URL: https://udspace.udel.edu/handle/19716/29071

        This research used resources of the Oak Ridge Leadership Computing Facility at the Oak Ridge National Laboratory, which is supported by the Office of Science of the U.S. Department of Energy under Contract No. DE-AC05-00OR22725.

%     \bibliography{myref}

\providecommand{\latin}[1]{#1}
\makeatletter
\providecommand{\doi}
  {\begingroup\let\do\@makeother\dospecials
  \catcode`\{=1 \catcode`\}=2 \doi@aux}
\providecommand{\doi@aux}[1]{\endgroup\texttt{#1}}
\makeatother
\providecommand*\mcitethebibliography{\thebibliography}
\csname @ifundefined\endcsname{endmcitethebibliography}
  {\let\endmcitethebibliography\endthebibliography}{}

\end{document}

\newpage

\tableofcontents

\clearpage

    \section{Unbiased simulations}
    \subsection{Classical molecular dynamics (MD) methods}
    
    Classical molecular dynamics simulations were performed in NAMD (v.\ 2.14 CUDA) \cite{Phillips2020} using the CHARMM36 force field with CMAP corrections (July 2022 release) \cite{Huang2013} along with the CHARMM general force field (CGenFF v.\ 4.6). \cite{Vanommeslaeghe2012} The CHARMM36 \texttt{toppar\_water\_ions.str} file was divided manually into topology and parameter segments to enable its use with \texttt{psfgen} (v.\ 2.0) \cite{Humphrey1996} and NAMD, respectively. Individual \texttt{.sdf} files of polyatomic ions were generated from SMILES strings using the NCI Chemical Identifier Resolver and converted to \texttt{.pdb} files using Open Babel (v.\ 3.1.0). \cite{OBoyle2011} Packmol (v.\ 18.169) \cite{Martinez2009} was used to generate initial coordinates for nine systems, each of which contained one anion species (acetate, ethylsulfonate or methylsulfate) and one cation species (guanidinium, imidazolium or methylammonium). For each system, 22 anion-cation pairs were solvated in 2470 TIP3P\cite{Jorgensen1983} water molecules to form simulation boxes of $\sim$ 42~{\AA} in length, corresponding to a salt concentration of $\sim$ 0.5~M. 
  
    The \texttt{psfgen} utility of VMD (v. 1.9.4a43) \cite{Humphrey1996} was used to generate a \texttt{.psf} structure file for each system, and the partial charges of ion atoms were scaled by a factor of 0.75 for simulations that employed the electronic continuum correction (ECC). \cite{Dijon2020, Leontyev2011, Leontyev2009} All simulations were performed in the $NpT$ ensemble with periodic boundary conditions and parameterized as designed for CGenFF in NAMD (i.e., by reading CHARMM36 protein, carbohydrate, lipid and nucleic acid parameter files prior to CGenFF and water parameters). Time steps of 2~fs were used and the temperature was maintained at 298~K using a Langevin thermostat with a 1~ps$^{-1}$ friction coefficient. A 1~bar pressure was maintained using a Langevin barostat with a 200~fs piston period and a 100~fs decay time and the SHAKE algorithm was used to constrain the geometry of water molecules.\cite{Ryckaert1977} Electrostatic interactions were computed using the particle mesh Ewald (PME) method \cite{Darden1993, Essmann1995} with a grid spacing of 0.5~{\AA} and nonbonded interactions were truncated beyond 9.5~{\AA}. Nonbonded interactions among bonded atoms were treated with a scaled 1-4 exclusion policy without modification of electrostatic interactions between 1-4 atom pairs. A 1~ns equilibration simulation was followed by 100~ns of configurational sampling and frames were saved every 4~ps.

    \subsection{Semiclassical MD methods}
    Semiclassical simulations were performed in OpenMM (v.\ 7.7.0) \cite{Eastman2017} using the AMOEBA (2018) polarizable force field. \cite{Shi2013} Simulation parameters were generated for ions from \texttt{.sdf} files using Poltype 2 (v.\ 2.3.1) \cite{Walker2022} and appended to the \texttt{amoebabio18.prm} force field file that may be installed with Tinker (v.\ 8.10). \cite{Rackers2018} This was converted to \texttt{.xml} format using OpenMM's  \texttt{processTinkerForceField.py} script without processing ``special pair'' AMOEBA van der Waals parameters, which are not relevant to the systems of interest. A residue entry for each ion was added manually to the resulting \texttt{.xml} file as well as OpenMM's \texttt{residues.xml} file and atom names were assigned according to the CGenFF topology convention. 

    Semiclassical simulations used 1~fs time steps and were performed in mixed precision using the same $NpT$ boundary conditions as classical simulations. The \texttt{LangevinMiddleIntegrator} method was used with a 10~ps$^{-1}$ friction coefficient along with a Monte Carlo barostat. The PME method was used with truncation of nonbonded interactions beyond 10~{\AA} and dipole induction was estimated using the mutual polarization method with a tolerance of $1 \times 10^{-5}$. A 0.2~ns equilibration simulation was followed by 25~ns of configurational sampling and frames were saved every 0.1~ps.

    \subsection{Definition of $R_{Shell}$ for computing $K_A$}
    Potential of mean force (PMF) profiles were computed from simulation trajectories and standard errors of the mean were estimated using the Friedberg-Cameron algorithm.\cite{Friedberg1970, Allen1986} The first local maximum in each PMF profile was identified visually and used as the separation threshold $R_{Shell}$ that delineates whether a given ion pair is in the associated state. Unlike the other PMF profiles, those of imidazolium systems that were simulated using CHARMM36 with the ECC did not exhibit a local maximum following ion-pair contact but they did exhibit an inflection near that separation distance. That point was denoted a ``comparable" $R_{Shell}$ and used to facilitate comparison with the other imidazolium models along with an ``apparent" $R_{Shell}$ value that corresponds to a solvent-separated complex. These values were used to estimate ion-pair association constants $K_{A}$ using Equation~1
    % HARD CODED
    with standard variance addition rules for error propagation.

    \subsection{Supplementary results from unbiased simulations}
    
    \begin{figure}[H]
    \begin{center}
        \includegraphics[width=\textwidth]{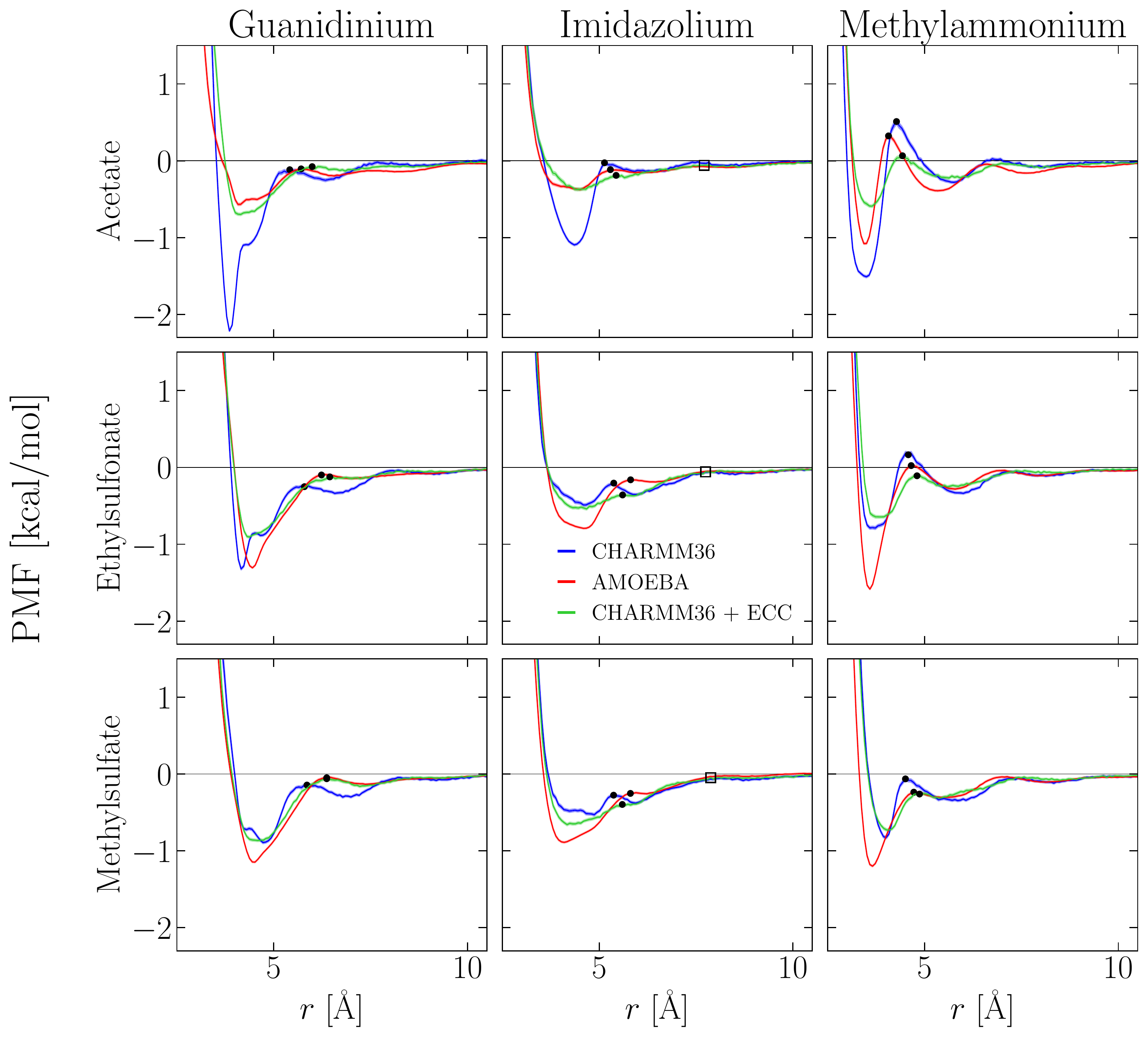}
        \caption{Potential of mean force (PMF) profiles from the interaction of cations (columns) with anions (rows) at a concentration of $\sim$ 0.5 M. The definition of the separation distance $r$ is described in Figure~1 and detailed in Table~\ref{tab:coordinates}. The black points denote the $R_{Shell}$ values that were used for estimating $K_{A}$ in Figure 2
        % HARD CODED
        and represent the comparable $R_{Shell}$ point for the CHARMM36 + ECC imidazolium systems, and black boxes represent the apparent $R_{Shell}$ for those systems. Line widths represent $2\times$ the standard error of the mean.}
        \label{sup-fig:pmfs}
    \end{center}
    \end{figure}

    \begin{figure}[H]
    \begin{center}
        \includegraphics[width=1\columnwidth]{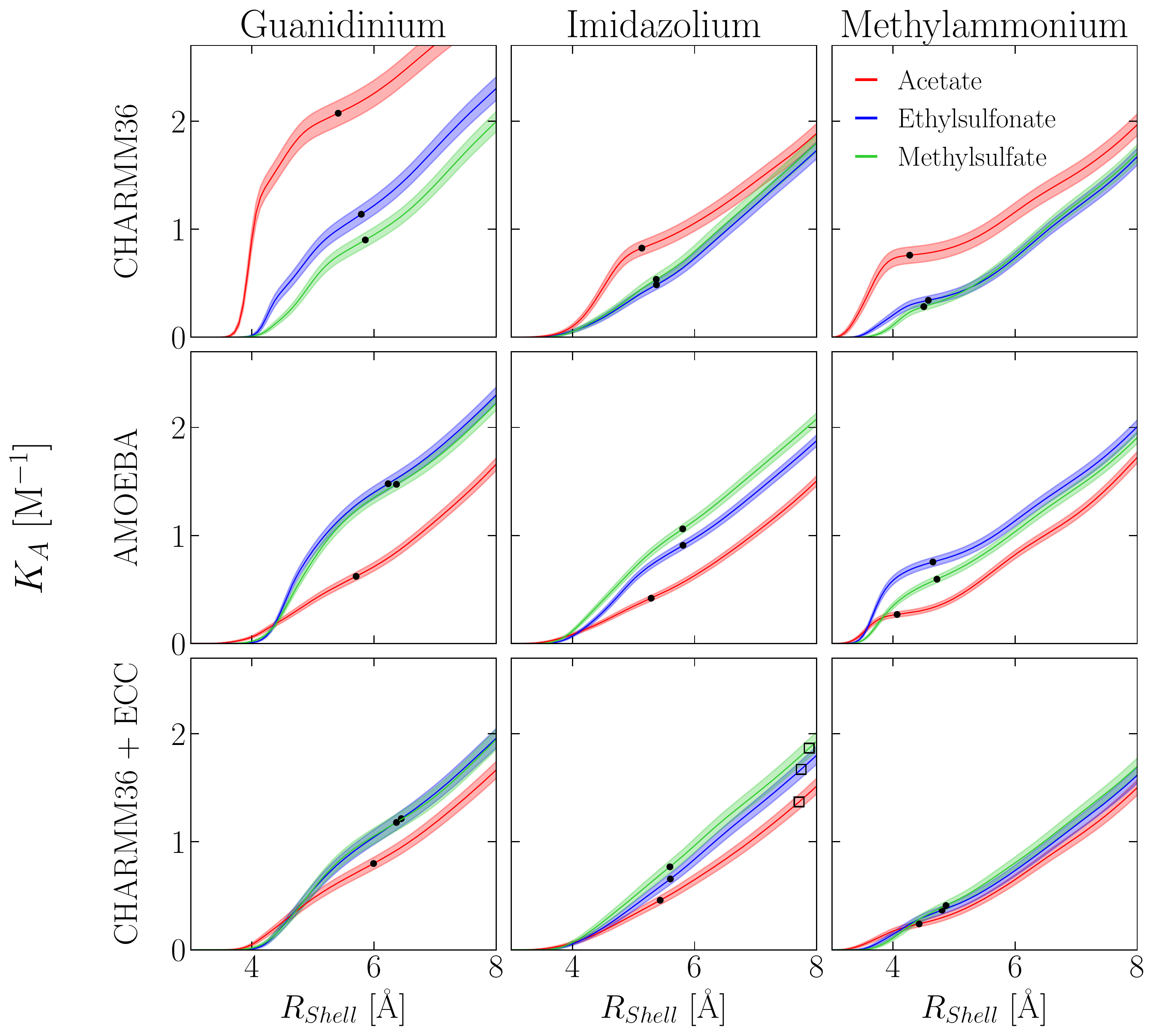}
        \caption{Dependence of the association constant $K_{A}$ on $R_{Shell}$. Here the rows correspond to the force fields and the lines correspond to the anions. The black points denote the $R_{Shell}$ values that were used for estimating $K_{A}$ in Figure 2
        % HARD CODED
        and represent the comparable $R_{Shell}$ points for the CHARMM36 + ECC imidazolium systems, and black boxes represent the apparent $R_{Shell}$ values for those systems. Shaded regions represent $2\times$ the standard error of the mean.}
        \label{fig:Ka_vs_Rshell}
    \end{center}
    \end{figure}

    \begin{adjustbox}{center, caption={$R_{Shell}$, $K_{A}$ and PMF minima of the anion-cation pairs as simulated in the different force fields. Uncertainty values represent the standard error of the mean.}, float=table}
    \label{tab:Ka_data}
    \centering
    \renewcommand{\arraystretch}{0.68}
    \begin{tabular}{l | l | l | c | c | c }
    Force field    & Anion          & Cation          & $R_{Shell}$ [\AA] & $K_{A}$ [M$^{-1}$] & PMF min [kcal/mol] \\
    \hline
    CHARMM36       & Acetate        & Guanidinium     & 5.41  & $2.073 \pm 0.059$ & $-2.215 \pm 0.003$ \\
    CHARMM36       & Ethylsulfonate & Guanidinium     & 5.79  & $1.138 \pm 0.042$ & $-1.321 \pm 0.007$ \\
    CHARMM36       & Methylsulfate  & Guanidinium     & 5.86  & $0.900 \pm 0.034$ & $-0.892 \pm 0.007$ \\
    CHARMM36       & Acetate        & Imidazolium     & 5.14  & $0.825 \pm 0.033$ & $-1.094 \pm 0.006$ \\
    CHARMM36       & Ethylsulfonate & Imidazolium     & 5.38  & $0.485 \pm 0.022$ & $-0.489 \pm 0.008$ \\
    CHARMM36       & Methylsulfate  & Imidazolium     & 5.37  & $0.535 \pm 0.024$ & $-0.530 \pm 0.008$ \\
    CHARMM36       & Acetate        & Methylammonium  & 4.27  & $0.759 \pm 0.038$ & $-1.509 \pm 0.007$ \\
    CHARMM36       & Ethylsulfonate & Methylammonium  & 4.58  & $0.342 \pm 0.020$ & $-0.789 \pm 0.009$ \\
    CHARMM36       & Methylsulfate  & Methylammonium  & 4.51  & $0.283 \pm 0.018$ & $-0.828 \pm 0.008$ \\ \hline
    AMOEBA         & Acetate        & Guanidinium     & 5.71  & $0.622 \pm 0.019$ & $-0.567 \pm 0.008$ \\
    AMOEBA         & Ethylsulfonate & Guanidinium     & 6.23  & $1.479 \pm 0.030$ & $-1.307 \pm 0.004$ \\
    AMOEBA         & Methylsulfate  & Guanidinium     & 6.37  & $1.473 \pm 0.027$ & $-1.147 \pm 0.004$ \\
    AMOEBA         & Acetate        & Imidazolium     & 5.29  & $0.419 \pm 0.013$ & $-0.374 \pm 0.006$ \\
    AMOEBA         & Ethylsulfonate & Imidazolium     & 5.81  & $0.909 \pm 0.019$ & $-0.793 \pm 0.004$ \\
    AMOEBA         & Methylsulfate  & Imidazolium     & 5.81  & $1.061 \pm 0.021$ & $-0.889 \pm 0.004$ \\
    AMOEBA         & Acetate        & Methylammonium  & 4.07  & $0.269 \pm 0.012$ & $-1.077 \pm 0.005$ \\
    AMOEBA         & Ethylsulfonate & Methylammonium  & 4.65  & $0.755 \pm 0.021$ & $-1.585 \pm 0.004$ \\
    AMOEBA         & Methylsulfate  & Methylammonium  & 4.72  & $0.595 \pm 0.017$ & $-1.200 \pm 0.004$ \\ \hline
    CHARMM36 + ECC & Acetate        & Guanidinium     & 5.99  & $0.799 \pm 0.028$ & $-0.698 \pm 0.008$ \\
    CHARMM36 + ECC & Ethylsulfonate & Guanidinium     & 6.45  & $1.215 \pm 0.037$ & $-0.906 \pm 0.007$ \\
    CHARMM36 + ECC & Methylsulfate  & Guanidinium     & 6.37  & $1.179 \pm 0.038$ & $-0.865 \pm 0.007$ \\
    CHARMM36 + ECC & Acetate        & Imidazolium     & 5.44  & $0.459 \pm 0.020$ & $-0.372 \pm 0.009$ \\
    & & (comp. $R_{Shell}$) & & & \\
    CHARMM36 + ECC & Ethylsulfonate & Imidazolium     & 5.60 & $0.656 \pm 0.026$ & $-0.540 \pm 0.007$ \\
    & & (comp. $R_{Shell}$) & & & \\
    CHARMM36 + ECC & Methylsulfate  & Imidazolium     & 5.60 & $0.768 \pm 0.029$ & $-0.658 \pm 0.008$ \\
    & & (comp. $R_{Shell}$) & & & \\
    CHARMM36 + ECC & Acetate        & Imidazolium     & 7.71 & $1.369 \pm 0.035$ & $-0.372 \pm 0.009$ \\
    & & (app. $R_{Shell}$) & & & \\
    CHARMM36 + ECC & Ethylsulfonate & Imidazolium     & 7.75 & $1.671 \pm 0.040$ & $-0.540 \pm 0.007$ \\
    & & (app. $R_{Shell}$) & & & \\
    CHARMM36 + ECC & Methylsulfate  & Imidazolium     & 7.88 & $1.867 \pm 0.043$ & $-0.658 \pm 0.008$ \\
    & & (app. $R_{Shell}$) & & & \\
    CHARMM36 + ECC & Acetate        & Methylammonium  & 4.43  & $0.241 \pm 0.014$ & $-0.590 \pm 0.009$ \\
    CHARMM36 + ECC & Ethylsulfonate & Methylammonium  & 4.80  & $0.367 \pm 0.018$ & $-0.644 \pm 0.008$ \\
    CHARMM36 + ECC & Methylsulfate  & Methylammonium  & 4.87  & $0.410 \pm 0.020$ & $-0.730 \pm 0.008$
    \end{tabular}
    \end{adjustbox}
    \clearpage
    
    \begin{figure}[H]
    \begin{center}
        \includegraphics[width=0.9\columnwidth]{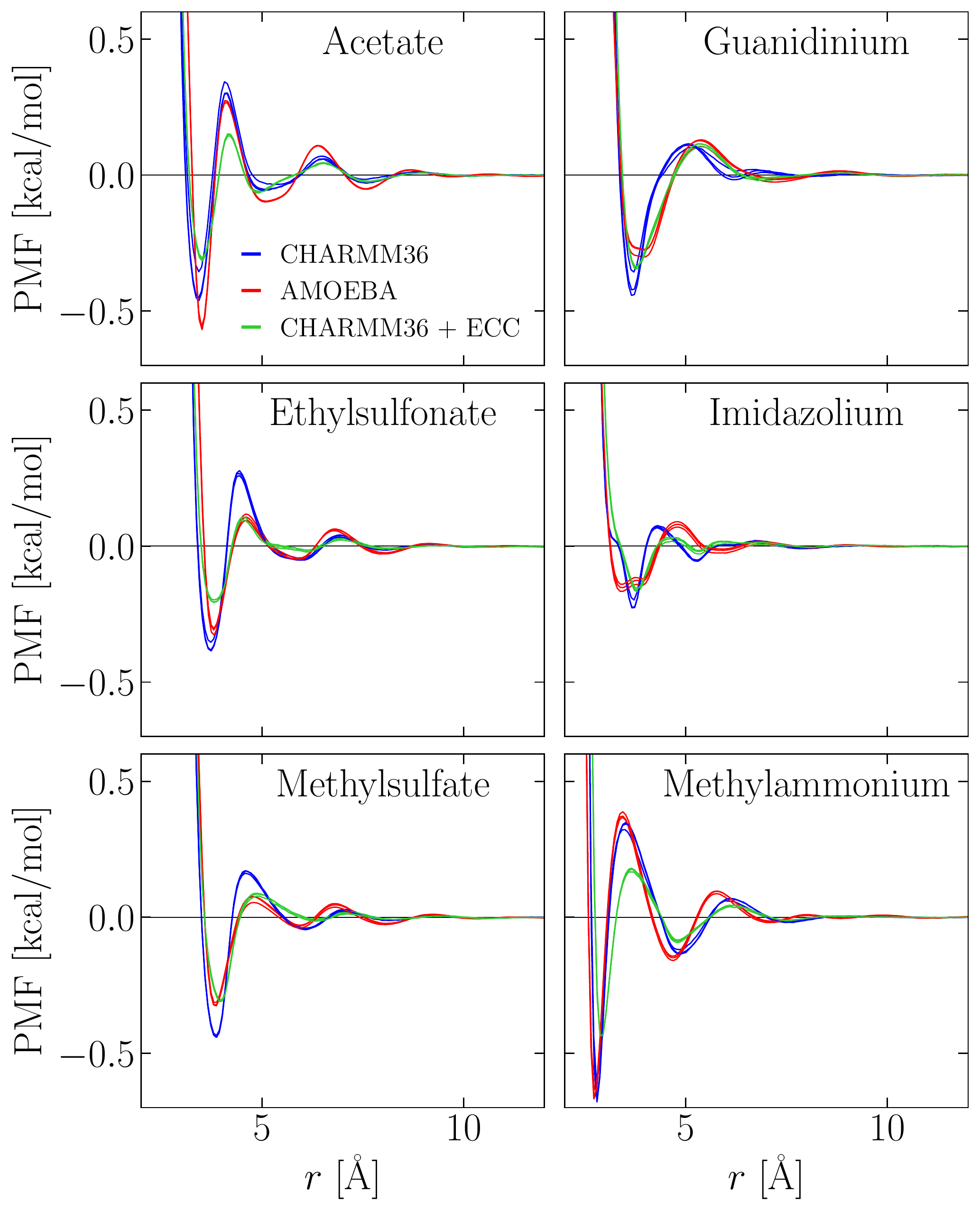}
        \caption{Overlay of ion-water potential of mean force (PMF) profiles from simulations with the three counterions, where $r$ is defined in this plot as the separation distance between the water oxygen atom and the ``central" ion atom that is highlighted in Figure 1 with a grey or blue circle (for anions and cations, respectively).}
        \label{fig:water_pmfs}
    \end{center}
    \end{figure}

    \begin{figure}[H]
    \begin{center}
        \includegraphics[width=0.6\columnwidth]{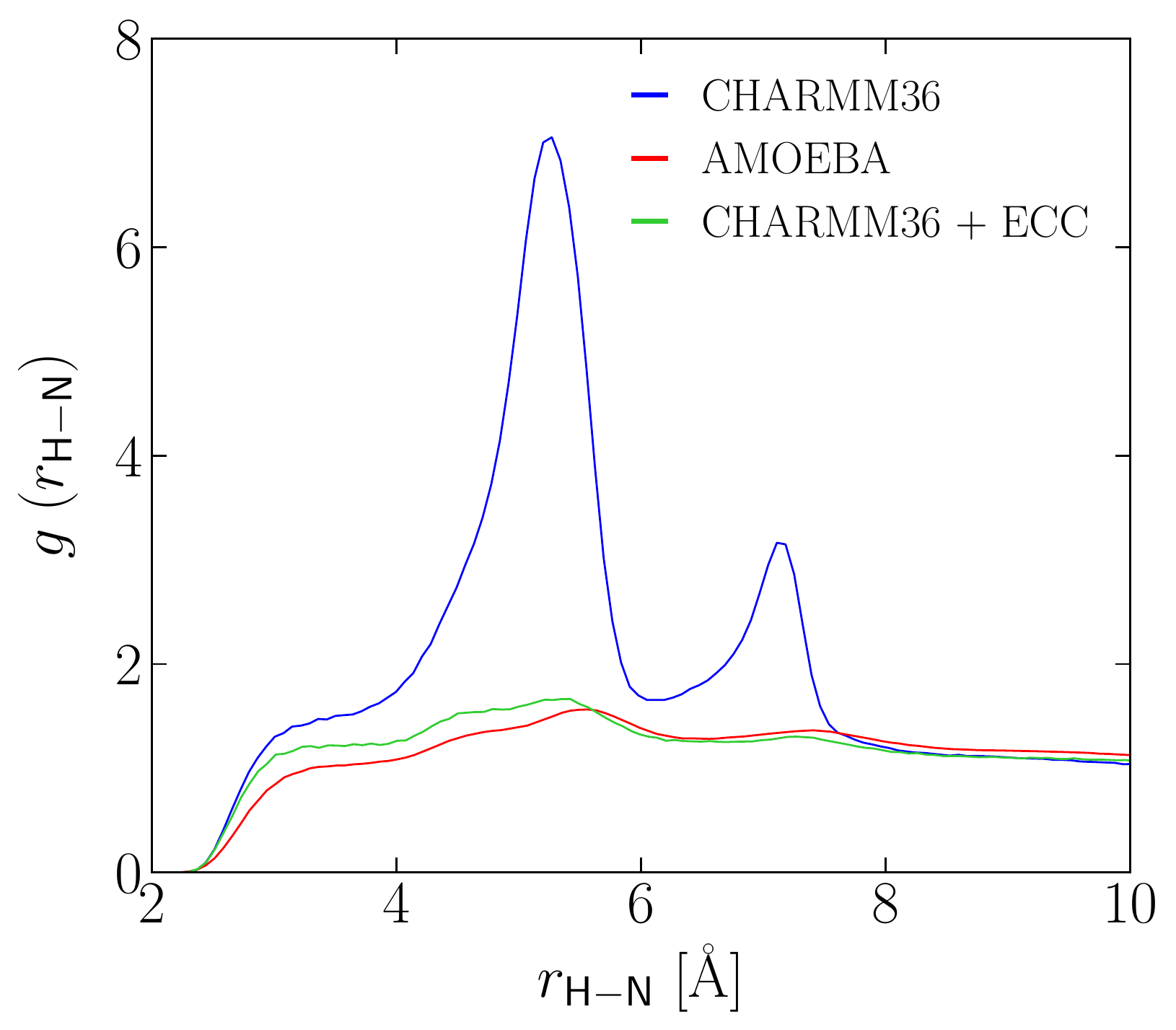}
        \caption{Radial distribution function between acetate hydrogen and guanidinium nitrogen atoms (separation distance $r_{\rm{H-N}}$) for comparison with Mason et al.\cite{Mason2019a}}
        \label{fig:gr_out_of_plane}
    \end{center}
    \end{figure}

    % \clearpage
    \section{\emph{In vacuo} energy analyses}
    
    \subsection{Methods}
    To estimate the underlying balance of attractive and repulsive forces that manifest as ion-pair association, the \emph{in vacuo} interaction energy between every anion-cation pair was computed for all configurations in the simulation trajectories. The \texttt{pairInteraction} module in NAMD was used to analyze both the CHARMM36 and AMOEBA trajectories using the CHARMM36 force field and electrostatic contributions were computed directly (i.e., without PME) with truncation at half the box length to obviate the need for Ewald self-interaction corrections. To enable a comparable analysis of the AMOEBA trajectories with the AMOEBA force field, \texttt{.arc} files of individual ions and anion-cation pairs were generated and used as input to the Tinker potential energy program \texttt{analyze}, with the potential energies of individual ions subtracted from the respective ion pairs. The contributions to the total \emph{in vacuo} interaction energy $U_{Total}$ of ion pairs were then binned according to the separation distance $r$ and ensemble-averaged. The solvent opposition to ion-pair association was inferred as $W_{Solv}~=~w~-~U_{Total}$, where $w$ represents the PMF.

    \subsection{Cross-FF analysis results}
    
    \begin{figure}[H]
    \begin{center}
        \includegraphics[width=1\columnwidth]{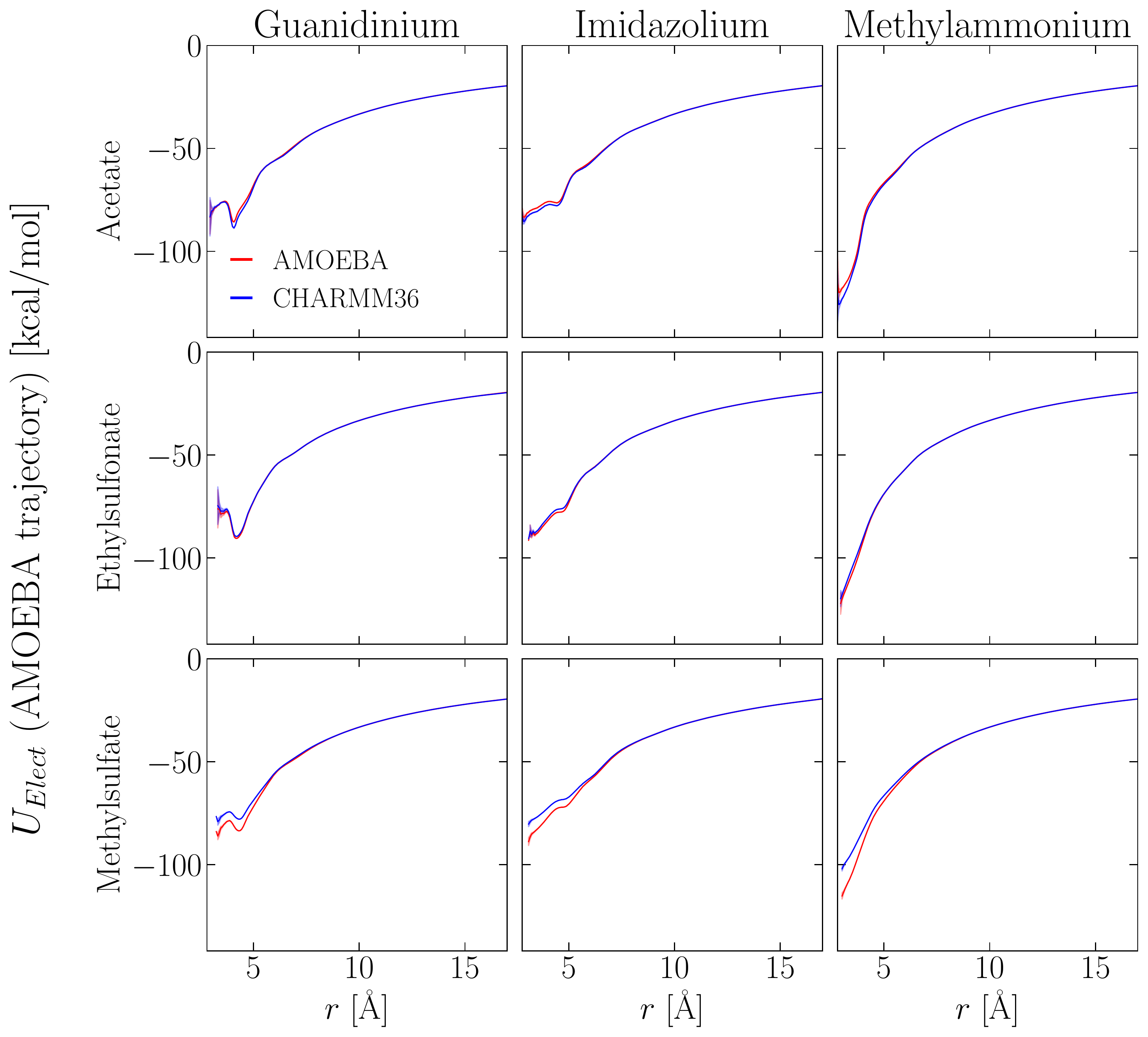}
        \caption{Permanent electrostatic contributions ($U_{Elect}$) to the \emph{in vacuo} interaction energy between anion-cation pairs in configurations from the AMOEBA simulation trajectory when retrospectively analyzed using the AMOEBA and CHARMM36 force fields. Shaded regions represent $2\times$ the standard error of the mean.}
        \label{fig:cross_ff_elect}
    \end{center}
    \end{figure}

    \begin{figure}[H]
    \begin{center}
        \includegraphics[width=1\columnwidth]{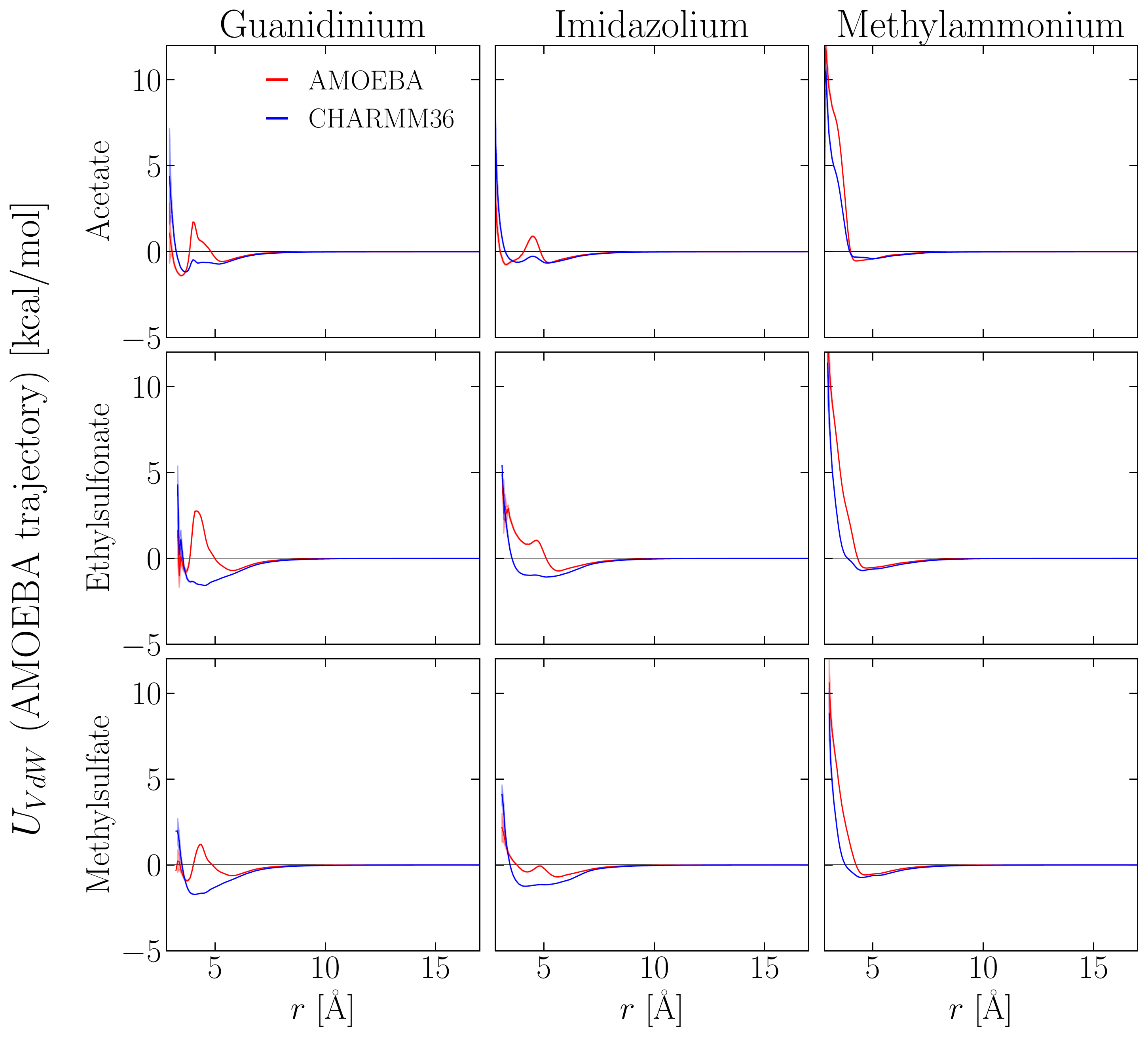}
        \caption{Van der Waals contributions ($U_{VdW}$) to the \emph{in vacuo} interaction energy between anion-cation pairs in configurations from the AMOEBA simulation trajectory when retrospectively analyzed using the AMOEBA and CHARMM36 force fields. Shaded regions represent $2\times$ the standard error of the mean.}
        \label{fig:cross_ff_vdw}
    \end{center}
    \end{figure}

    \begin{figure}[H]
    \begin{center}
        \includegraphics[width=1\columnwidth]{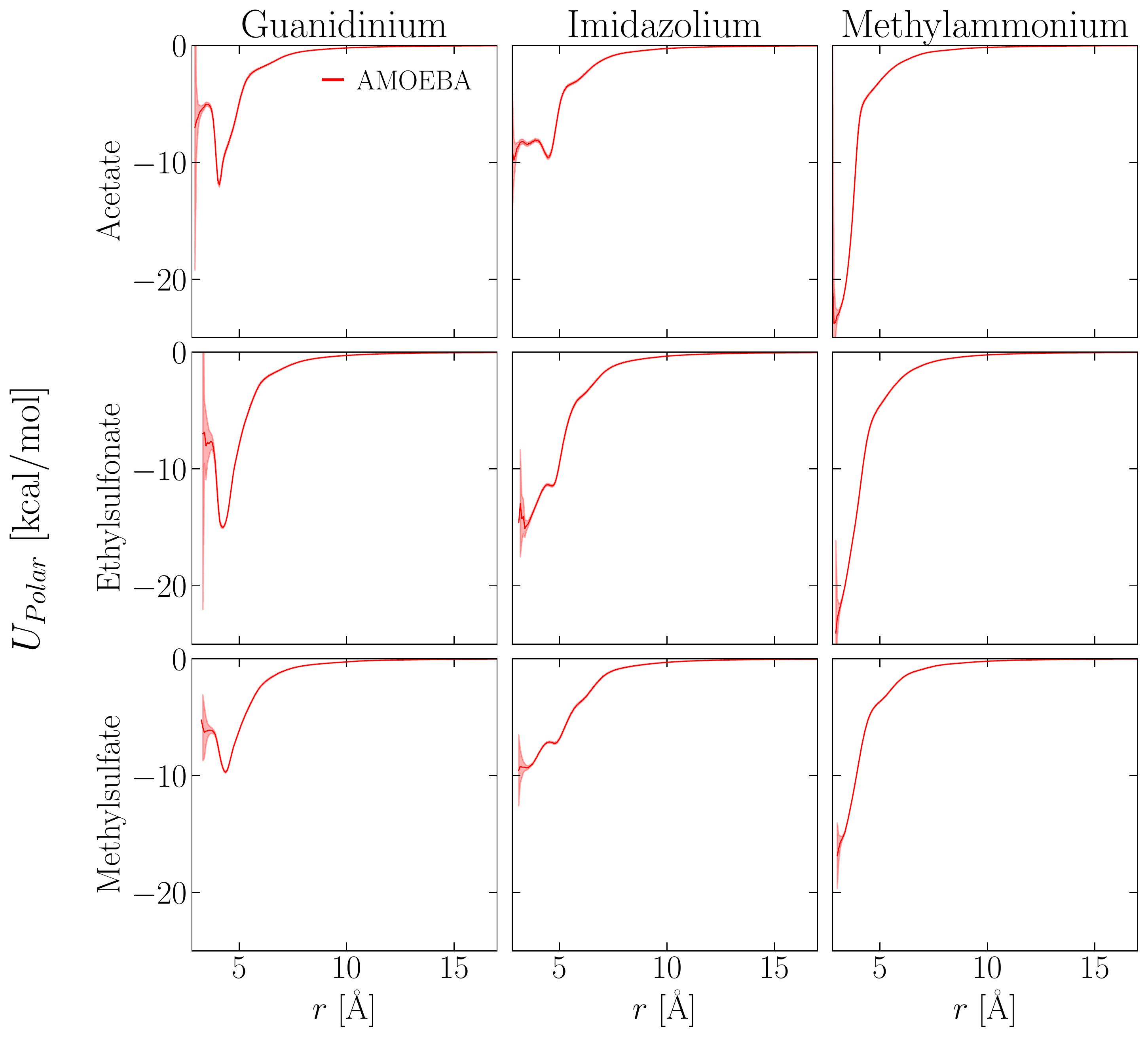}
        \caption{Estimated contribution of polarization ($U_{Polar}$) to the \emph{in vacuo} interaction energy between anion-cation pairs in configurations from the AMOEBA simulation trajectory when retrospectively analyzed using the AMOEBA force field. Shaded regions represent $2\times$ the standard error of the mean. (There are no polarization contributions in CHARMM36, so $U_{Polar}$ is identically zero for that force field.)}
        \label{fig:energy_conts_polar}
    \end{center}
    \end{figure}

    \begin{figure}[H]
    \begin{center}
        \includegraphics[width=1\columnwidth]{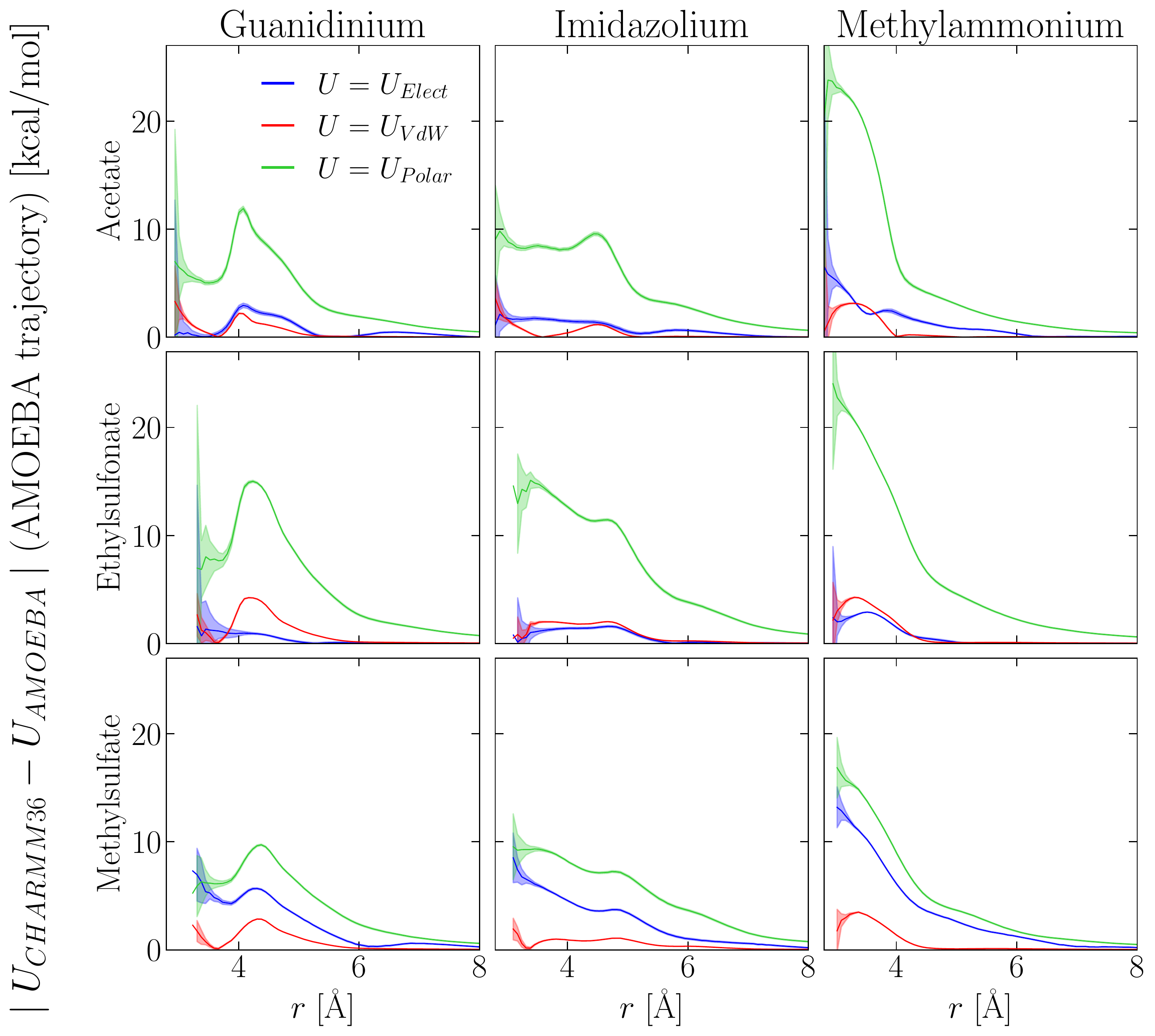}
        \caption{Magnitude of the differences between \emph{in vacuo} interaction energy contributions for anion-cation pairs in configurations from the AMOEBA simulation trajectory when retrospectively analyzed using the AMOEBA and CHARMM36 force fields. Notice that polarization stabilizes ion-pairing interactions \emph{in vacuo}. Shaded regions represent $2\times$ the standard error of the mean. (In the computation of differences in polarization energies, $U_{CHARMM36} = 0$.) }
        \label{fig:cross_ff_diff}
    \end{center}
    \end{figure}

    \subsection{PMF decomposition results}

    \begin{figure}[H]
    \begin{center}
        \includegraphics[width=1\columnwidth]{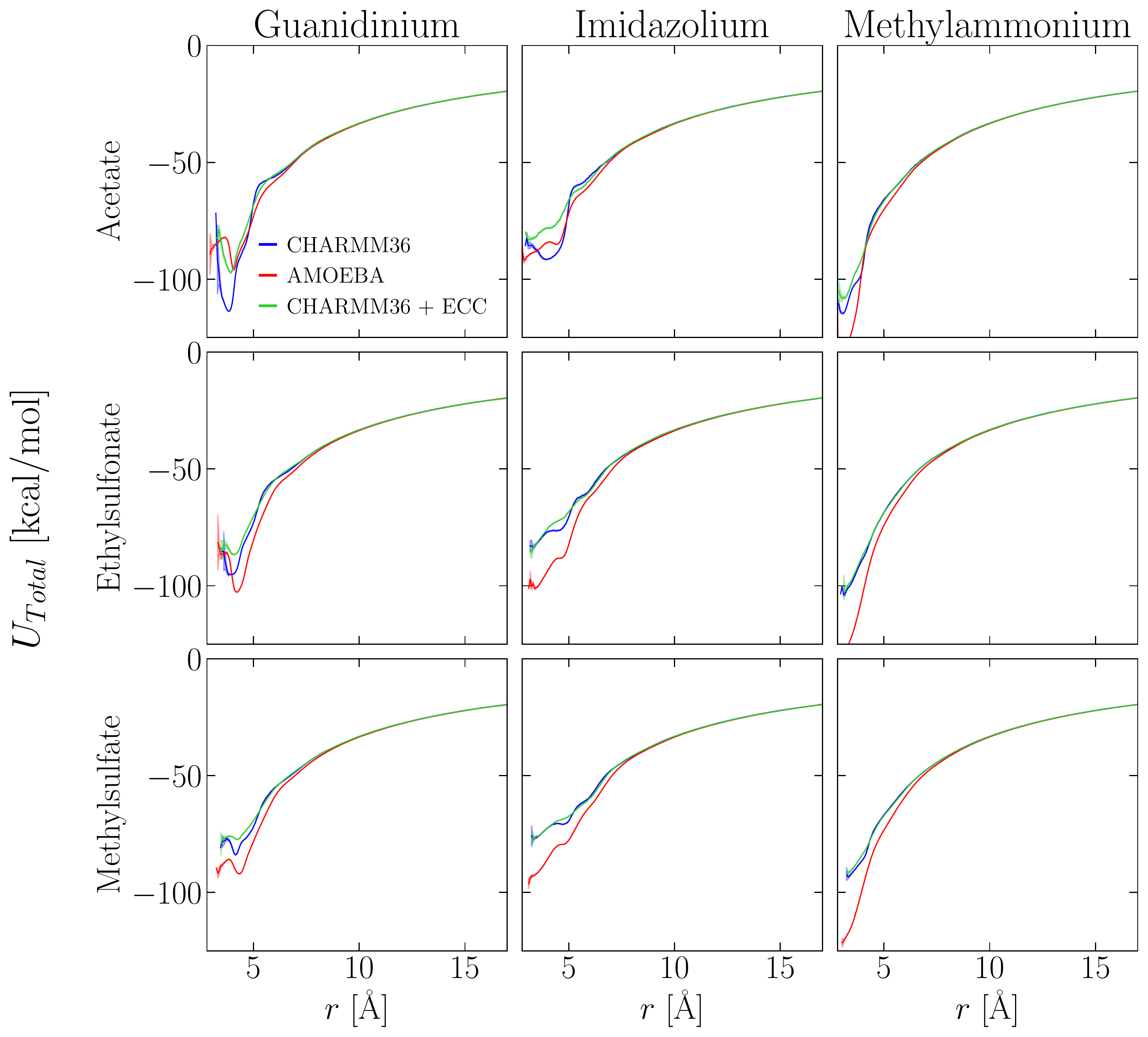}
        \caption{Total \emph{in vacuo} interaction energy between anion-cation pairs, computed as $U_{Total} = U_{Elect} + U_{VdW} + U_{Polar}$, where $U_{Elect}$ has been scaled up by $1/0.75^2$ for the CHARMM36 + ECC results. Shaded regions represent $2\times$ the standard error of the mean.}
        \label{fig:energy_conts_direct}
    \end{center}
    \end{figure}

    \begin{figure}[H]
    \begin{center}
        \includegraphics[width=1\columnwidth]{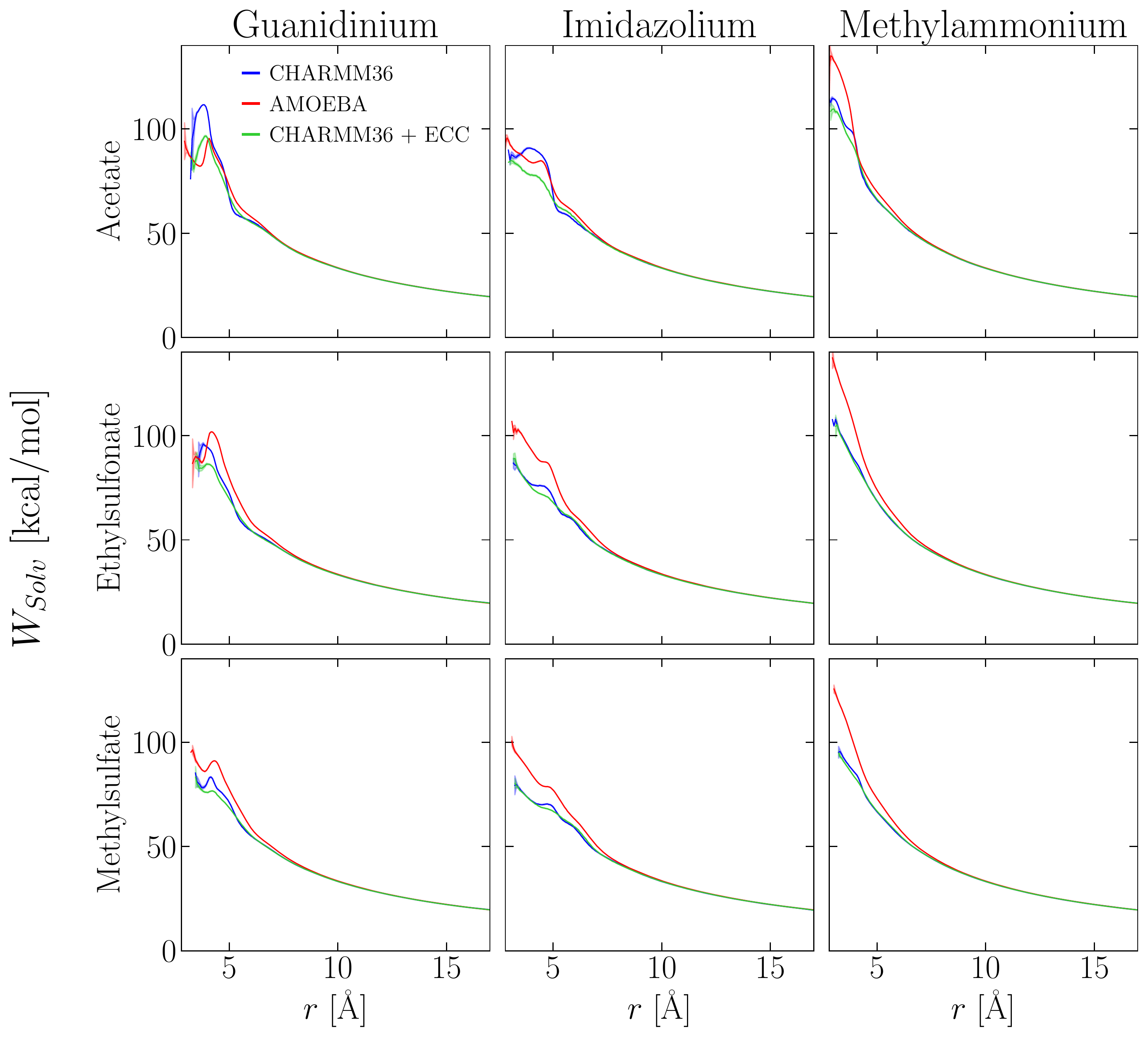}
        \caption{Estimated solvent opposition to ion-pair association, inferred as $W_{Solv}~=~\text{PMF}~-~U_{Total}$. $U_{Elect}$ was scaled up by $1/0.75^2$ in the computation of $U_{Total}$ for the CHARMM36 + ECC results. Shaded regions represent $2\times$ the standard error of the mean.}
        \label{fig:energy_conts_w_solv}
    \end{center}
    \end{figure}

    \begin{figure}[H]
    \begin{center}
        \includegraphics[width=1\columnwidth]{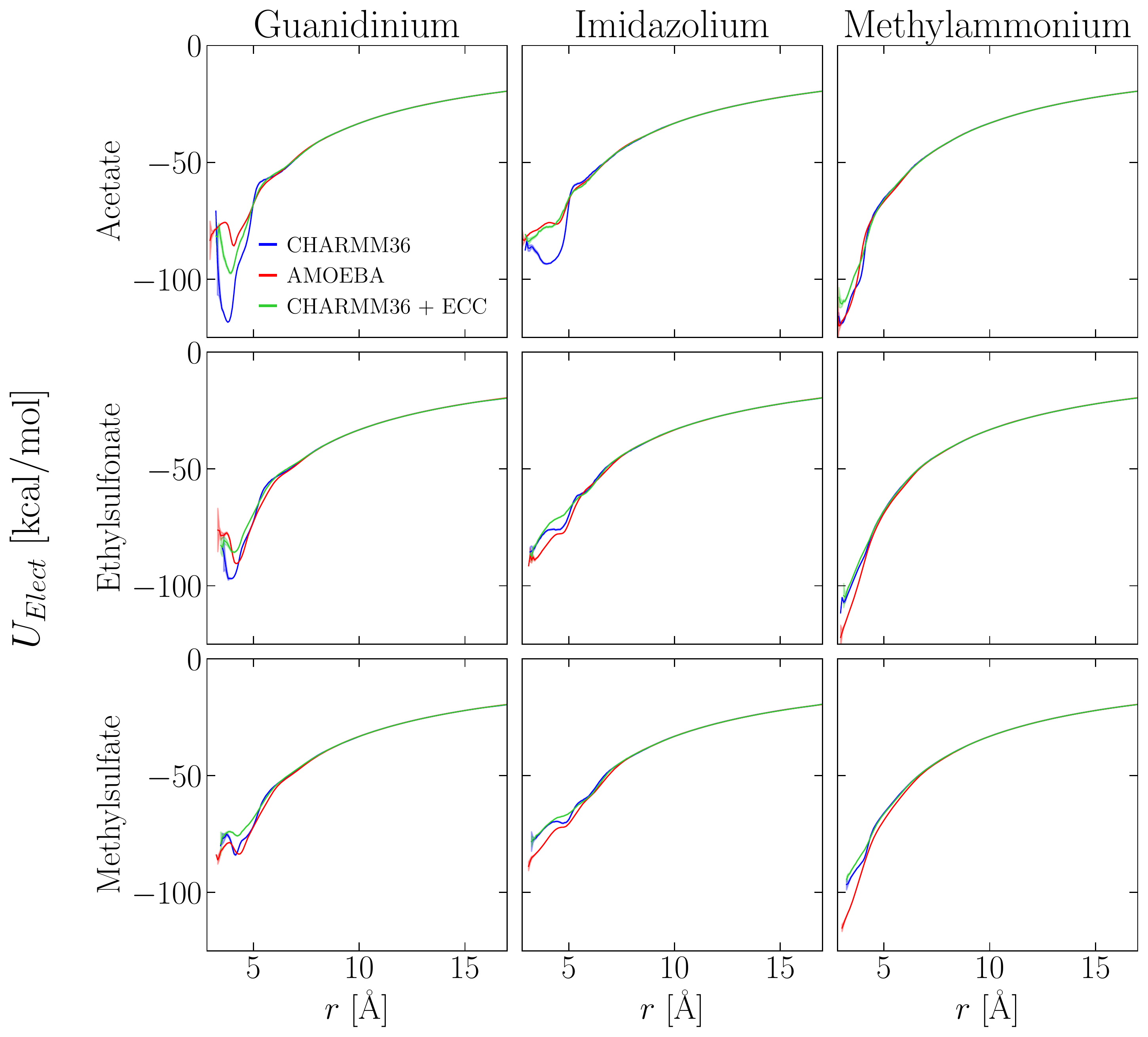}
        \caption{Permanent electrostatic contributions ($U_{Elect}$) to the \emph{in vacuo} interaction energy between anion-cation pairs in configurations from the different simulation trajectories when analyzed using the trajectory-generating force field. The CHARMM36 + ECC results have been scaled by a factor of $1/0.75^2$ to facilitate comparison. Shaded regions represent $2\times$ the standard error of the mean.}
        \label{fig:energy_conts_elect}
    \end{center}
    \end{figure}

    \begin{figure}[H]
    \begin{center}
        \includegraphics[width=1\columnwidth]{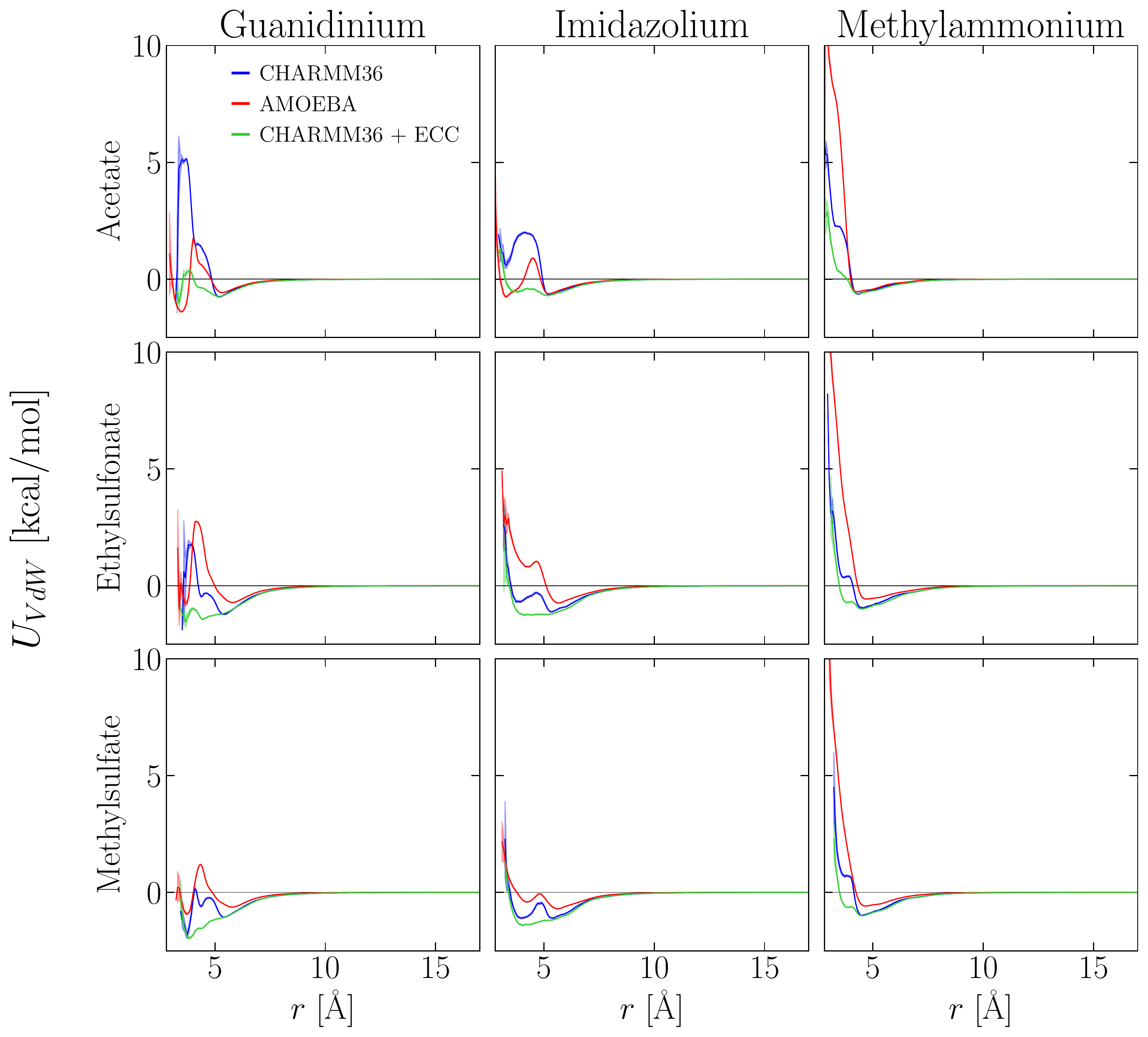}
        \caption{Van der Waals contributions ($U_{VdW}$) to the \emph{in vacuo} interaction energy between anion-cation pairs in configurations from the different simulation trajectories. Shaded regions represent $2\times$ the standard error of the mean.}
        \label{fig:energy_conts_vdw}
    \end{center}
    \end{figure}

    \begin{figure}[H]
    \begin{center}
        \includegraphics[width=0.9\columnwidth]{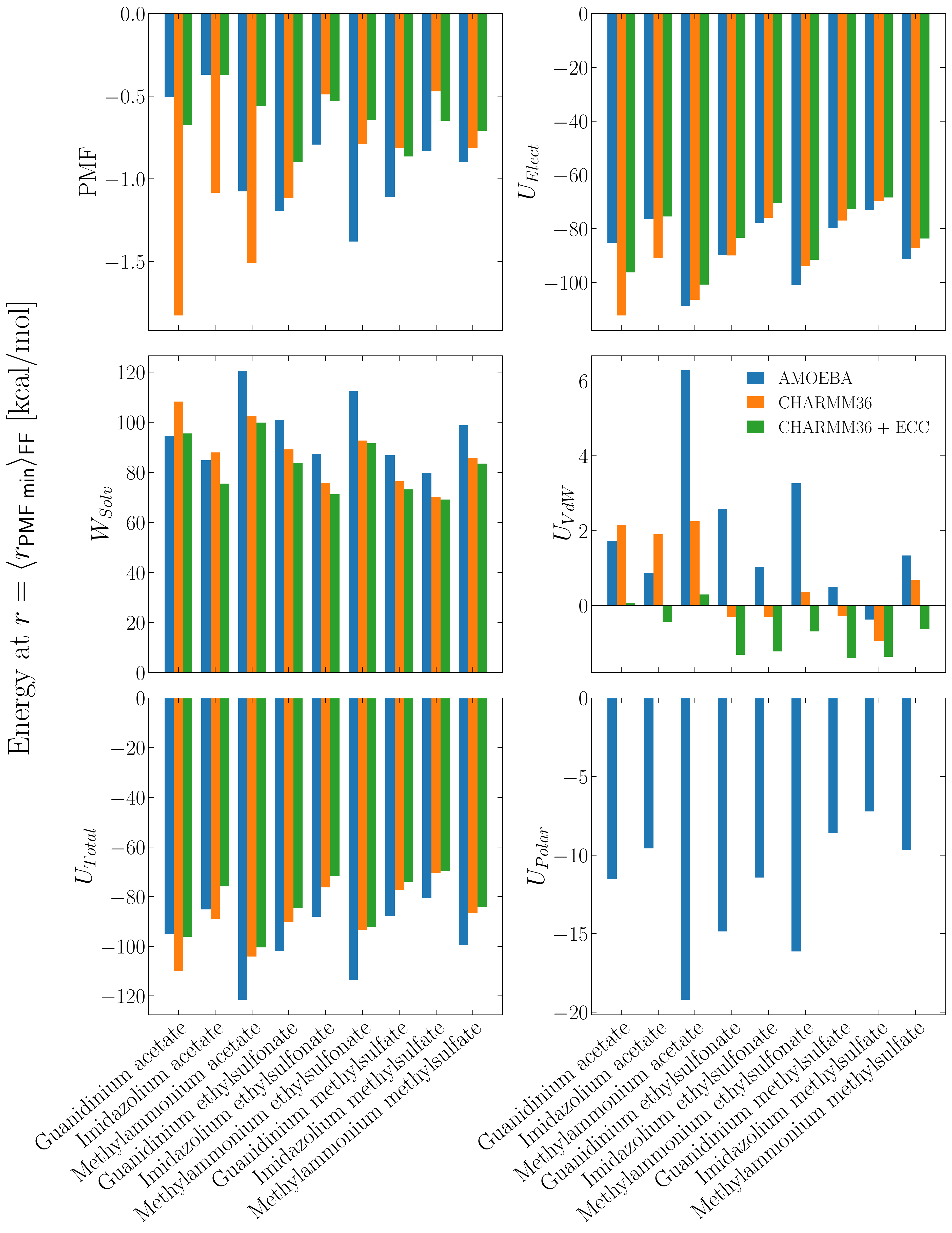}
        \caption{Interaction energy contributions near the PMF minimum (i.e., at the average $r_{PMF \; min}$ of the three force fields), where $U_{Elect}$ has been scaled up by $1/0.75^2$ for the CHARMM36 + ECC results in the presentation of $U_{Elect}$, $U_{Total}$ and $W_{Solv}$.}
        \label{fig:energy_cont_bar_plots}
    \end{center}
    \end{figure}

    \begin{landscape}
    \section{Maps of configurational sampling}
    \subsection{3D plots}
    
    \begin{figure}[H]
    \begin{center}
        \includegraphics[width=1.48\textwidth]{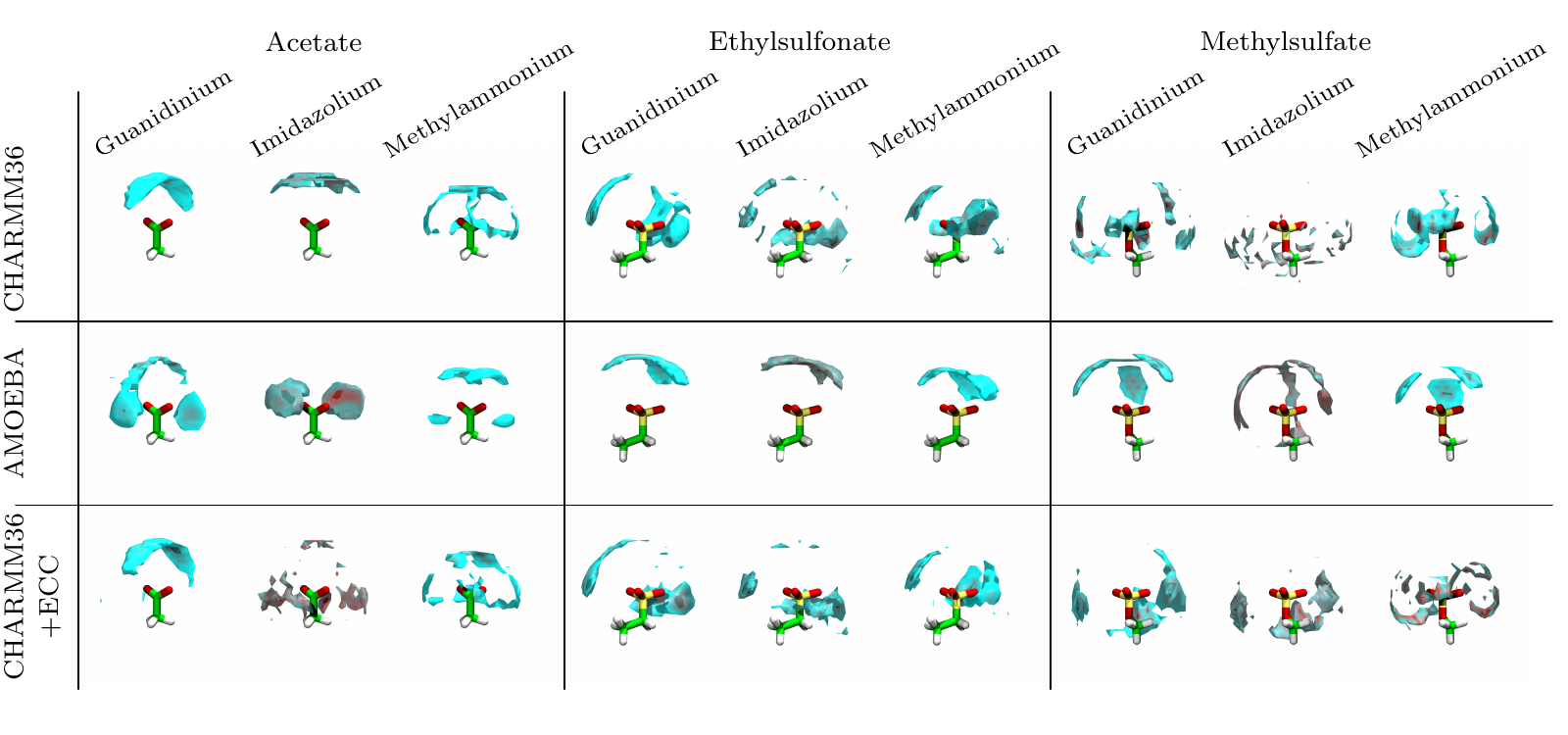}
        \caption{Spaces of most frequent cation sampling around anions. The shaded regions represent the top 1000 voxels in a cubic mesh (defined relative to the anion with a grid spacing of 0.5~\AA) where the sampling frequency was greatest for the ``central" cation atoms that are distinguished by blue circles in Figure~1. (Images were created using VMD.\cite{Humphrey1996})}
        \label{fig:3D_anions}
    \end{center}
    \end{figure}
    
    \begin{figure}[H]
    \begin{center}
        \includegraphics[width=1.48\textwidth]{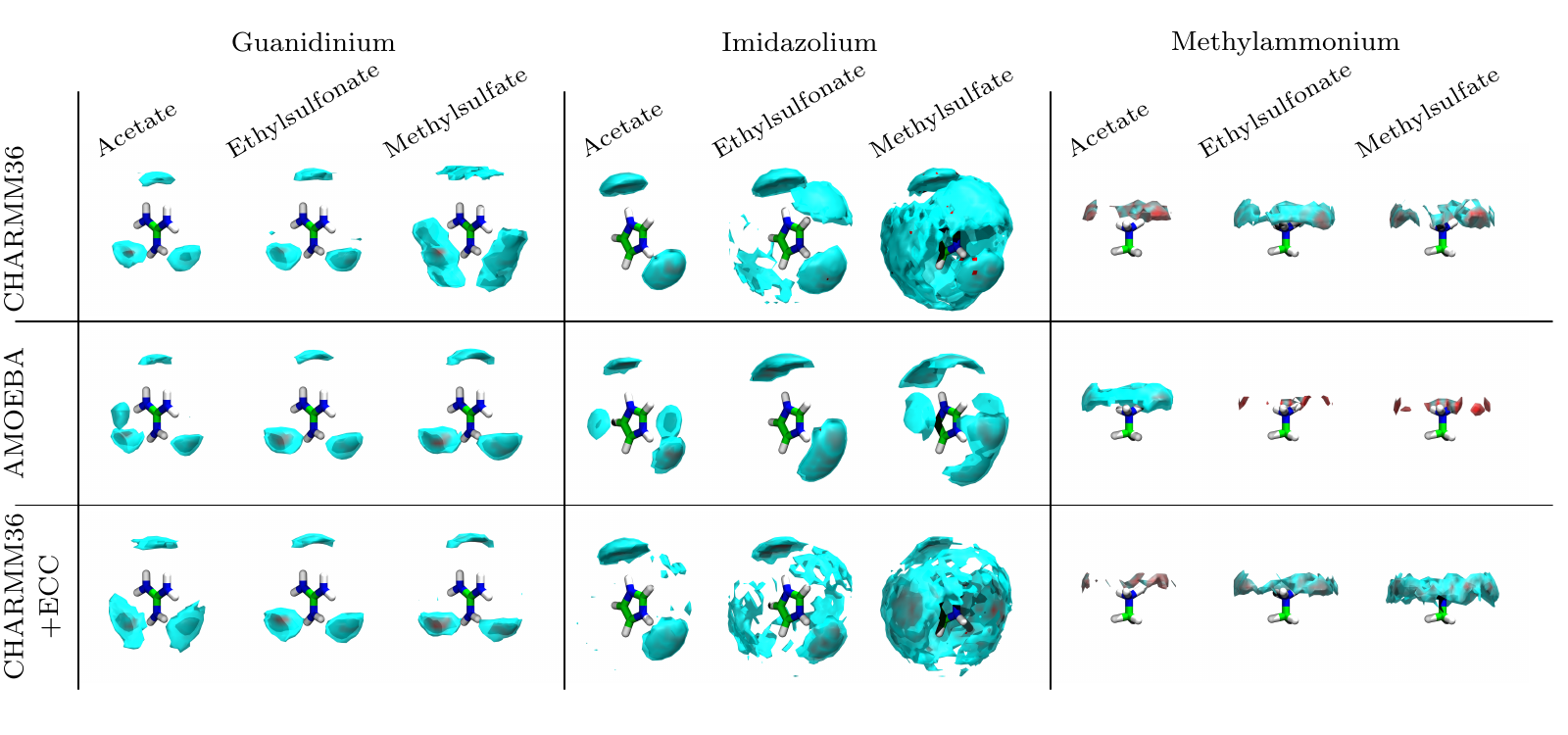}
        \caption{Spaces of most frequent anion sampling around cations. The shaded regions represent the top 1000 voxels in a cubic mesh (defined relative to the cation with a grid spacing of 0.5~\AA) where the sampling frequency was greatest for the ``central" anion atoms that are distinguished by grey circles in Figure~1. (Images were created using VMD.\cite{Humphrey1996})}
        \label{fig:3D_cations}
    \end{center}
    \end{figure}
    
    \end{landscape}

    \subsection{Coordinate definitions for heat maps}

\begin{figure}[H]
    \begin{center}
        \includegraphics[width=\columnwidth]{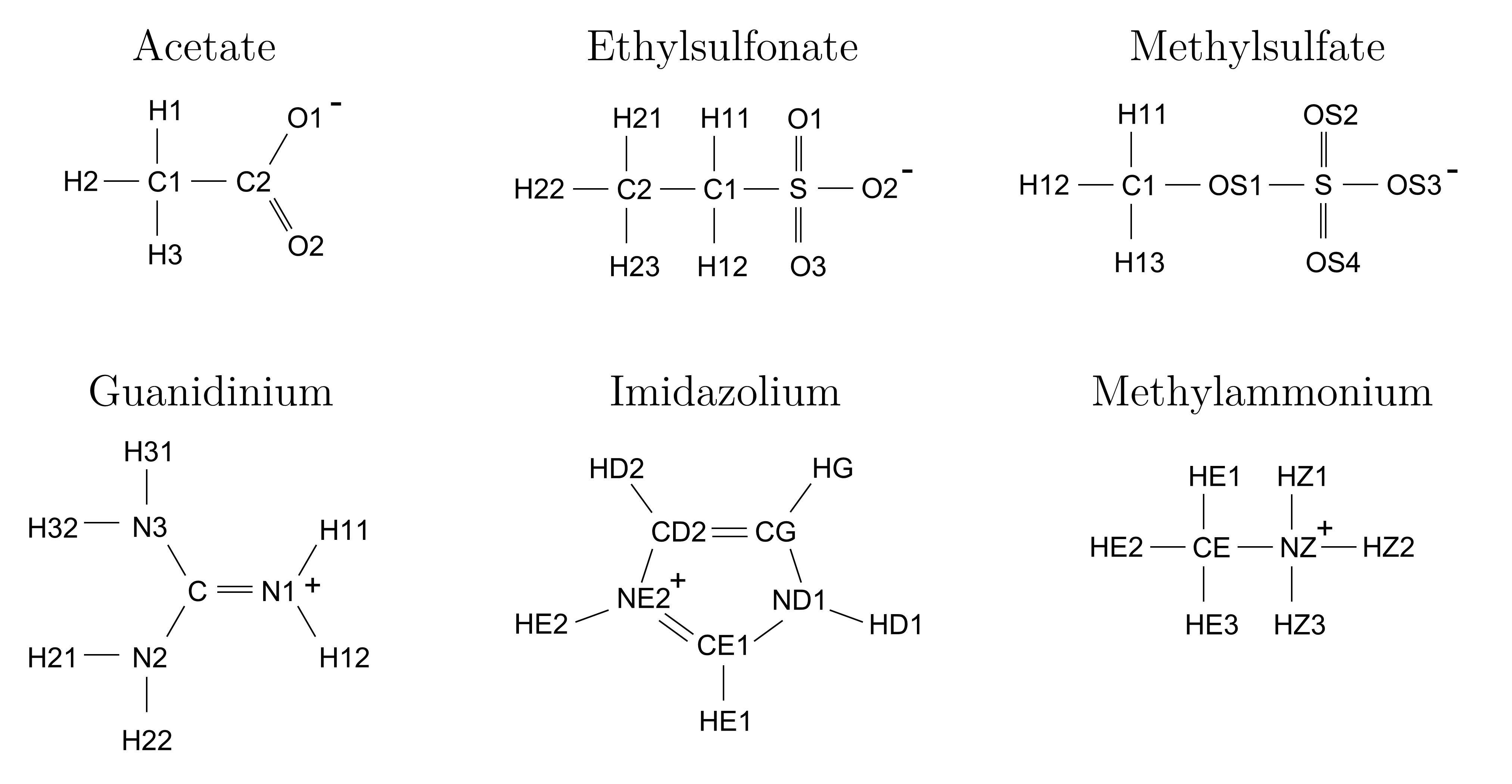}
        \caption{Ion structures with atom names from the CHARMM36 convention,\cite{Huang2013} where the first letter represents the element.}
        \label{fig:charmm_atom_names}
    \end{center}
    \end{figure}

    The separation distance $r = | \vec{r_1} |$ between a given anion-cation pair was defined using the central atoms that are distinguished in Figure 1, where grey and blue circles are used to distinguish the central atoms in anions and cations, respectively.
    % HARD CODED
    For clarity these atoms are also listed in Table~\ref{tab:coordinates} using the CHARMM36 atom names that are shown in Figure~\ref{fig:charmm_atom_names}. Angular coordinates $\theta$ and $\alpha$ were also defined to provide measures of the anion orientation relative to the cation and of the cation orientation relative to the anion, respectively. The definition of these coordinates depends on the ion pair and is explained in Table~\ref{tab:coordinates}.
    
    \begin{table}[H]
    \caption{Coordinate definitions for each anion-cation pair, where $A_{X}^{-}C_{Y}^{+}$ for example refers to the vector pointing from atom $X$ in the anion to atom $Y$ in the cation (cf. Figure~\ref{fig:charmm_atom_names} for atom names). The coordinate $r = | \vec{r_1} |$ represents the separation distance between the respective central atoms of the anion and the cation. The coordinate $\theta$ is computed as the angle between $\vec{r_1}$ and $\vec{r_2}$ and provides a measure of the anion orientation relative to the cation. The definition of $\alpha$ depends on the cation and provides a measure of the cation orientation relative to the anion.}
    \label{tab:coordinates}
    \centering
    \renewcommand{\arraystretch}{1.5}
    \begin{tabular}{>{\centering\arraybackslash}m{0.2\linewidth} | >{\centering\arraybackslash}m{0.17\linewidth} | >{\centering\arraybackslash}m{0.09\linewidth} | >{\centering\arraybackslash}m{0.09\linewidth} | >{\centering\arraybackslash}m{0.35\linewidth}}
    
    Cation         & Anion          & $\vec{r_1}$     & $\vec{r_2}$    & $\alpha$                                                                    \\ \hline
    Guanidinium    & Acetate        & $A_{C2}^{-}C_{C}^{+}$   & $A_{C2}^{-}A_{C1}^{-}$ & Angle formed by $\vec{r_1}$ and the plane defined by $C_{N1}^{+}C_{N2}^{+}C_{N3}^{+}$ \\
    Guanidinium    & Ethylsulfonate & $A_{S}^{-}C_{C}^{+}$    & $A_{S}^{-}A_{C1}^{-}$  & Angle formed by $\vec{r_1}$ and the plane defined by $C_{N1}^{+}C_{N2}^{+}C_{N3}^{+}$ \\
    Guanidinium    & Methylsulfate  & $A_{S}^{-}C_{C}^{+}$    & $A_{S}^{-}A_{OS1}^{-}$ & Angle formed by $\vec{r_1}$ and the plane defined by $C_{N1}^{+}C_{N2}^{+}C_{N3}^{+}$ \\ \hline
    Imidazolium    & Acetate        & $A_{C2}^{-}C_{CE1}^{+}$ & $A_{C2}^{-}A_{C1}^{-}$ & Angle formed by $\vec{r_1}$ and the plane defined by $C_{CG}^{+}C_{CD2}^{+}C_{CE1}^{+}$  \\
    Imidazolium    & Ethylsulfonate & $A_{S}^{-}C_{CE1}^{+}$  & $A_{S}^{-}A_{C1}^{-}$  & Angle formed by $\vec{r_1}$ and the plane defined by $C_{CG}^{+}C_{CD2}^{+}C_{CE1}^{+}$  \\
    Imidazolium    & Methylsulfate  & $A_{S}^{-}C_{CE1}^{+}$  & $A_{S}^{-}A_{OS1}^{-}$ & Angle formed by $\vec{r_1}$ and the plane defined by $C_{CG}^{+}C_{CD2}^{+}C_{CE1}^{+}$  \\ \hline
    Methylammonium & Acetate        & $A_{C2}^{-}C_{NZ}^{+}$  & $A_{C2}^{-}A_{C1}^{-}$ & Torsion angle formed by $C_{CE}^{+}C_{NZ}^{+}A_{C2}^{-}A_{C1}^{-}$ \\
    Methylammonium & Ethylsulfonate & $A_{S}^{-}C_{NZ}^{+}$   & $A_{S}^{-}A_{C1}^{-}$  & Torsion angle formed by $C_{CE}^{+}C_{NZ}^{+}A_{S}^{-}A_{C1}^{-}$ \\
    Methylammonium & Methylsulfate  & $A_{S}^{-}C_{NZ}^{+}$   & $A_{S}^{-}A_{OS1}^{-}$ & Torsion angle formed by $C_{CE}^{+}C_{NZ}^{+}A_{S}^{-}A_{OS1}^{-}$ 
    \end{tabular}
    \end{table}

    \newpage
    \subsection{Heat maps in the ($\theta$, $\alpha$) space for $r < R_{Shell}$}

\begin{figure}[H]
    \begin{center}
        \includegraphics[width=\textwidth]{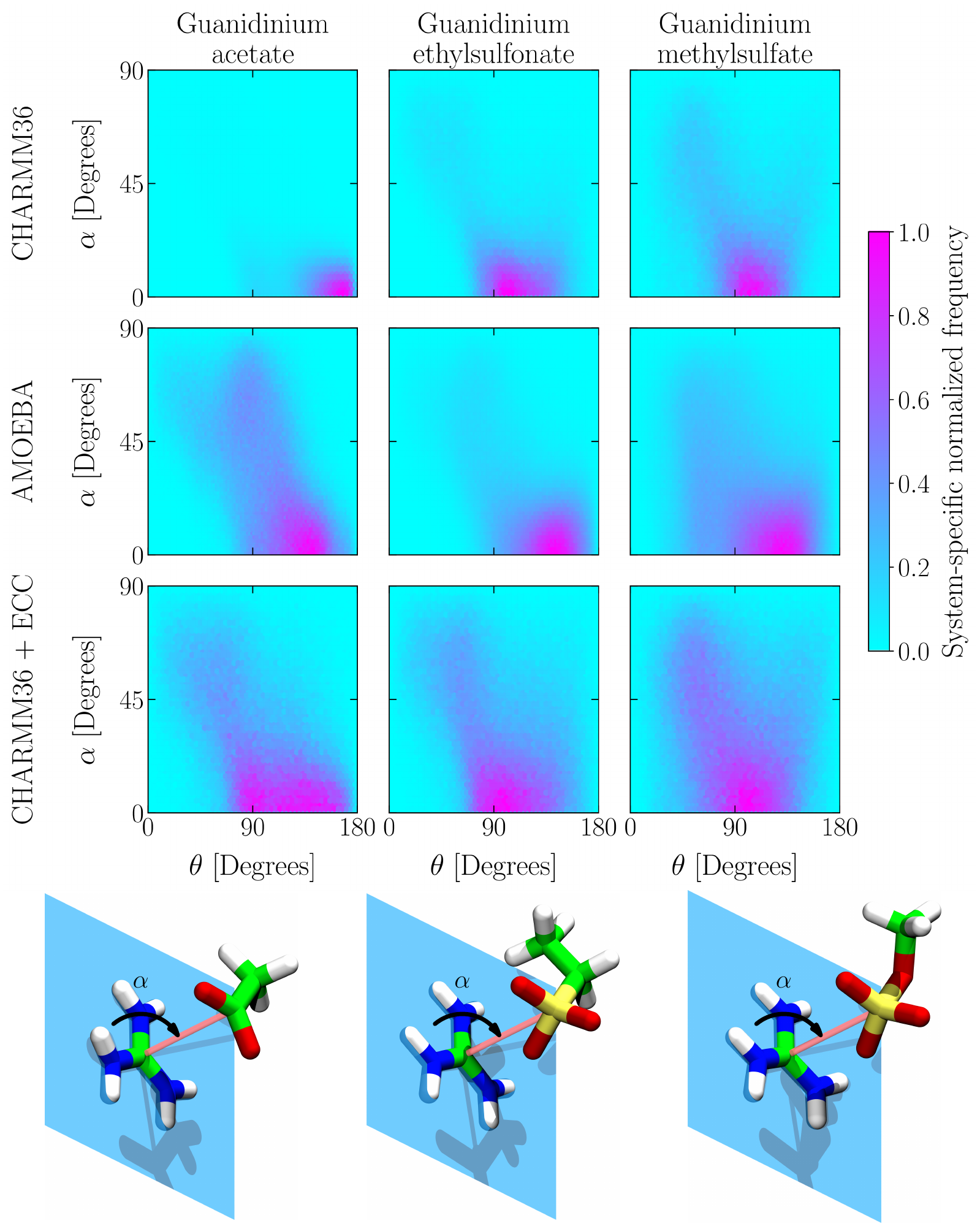}
        \caption{Heat maps of the sampling frequency of anion-guanidinium pairs (columns) with $r < R_{Shell}$ as observed in simulations using different force fields (rows). The normalization of sampling frequency is specific to each system. Refer to Table~\ref{tab:coordinates} for the definitions of $\theta$ and $\alpha$.}
        \label{fig:heatmap_guan}
    \end{center}
    \end{figure}

\begin{figure}[H]
    \begin{center}
        \includegraphics[width=\columnwidth]{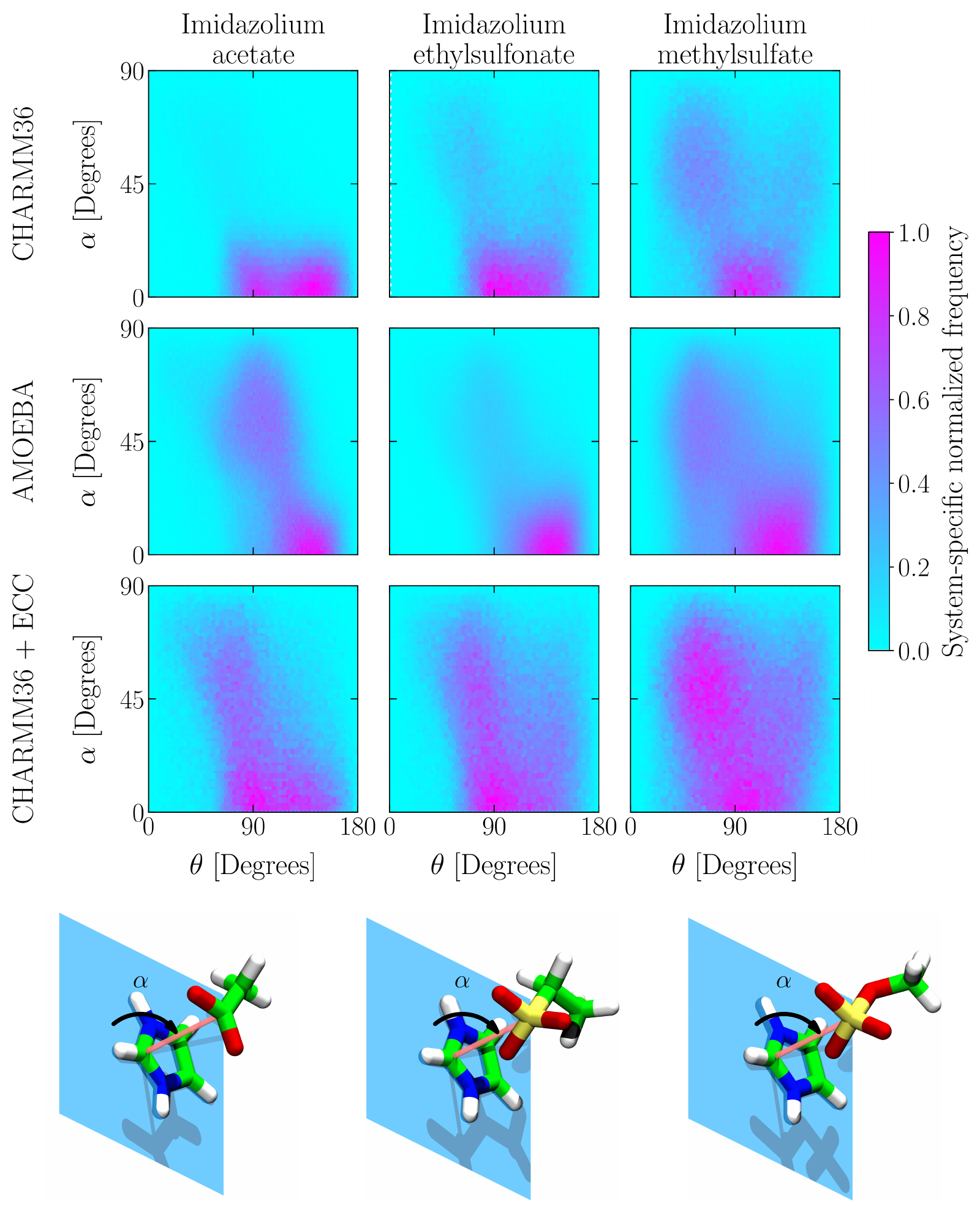}
        \caption{Heat maps of the sampling frequency of anion-imidazolium pairs (columns) with $r < R_{Shell}$ as observed in simulations using different force fields (rows). The normalization of sampling frequency is specific to each system. Refer to Table~\ref{tab:coordinates} for the definitions of $\theta$ and $\alpha$.}
        \label{fig:heatmap_imim}
    \end{center}
    \end{figure}

\begin{figure}[H]
    \begin{center}
        \includegraphics[width=\columnwidth]{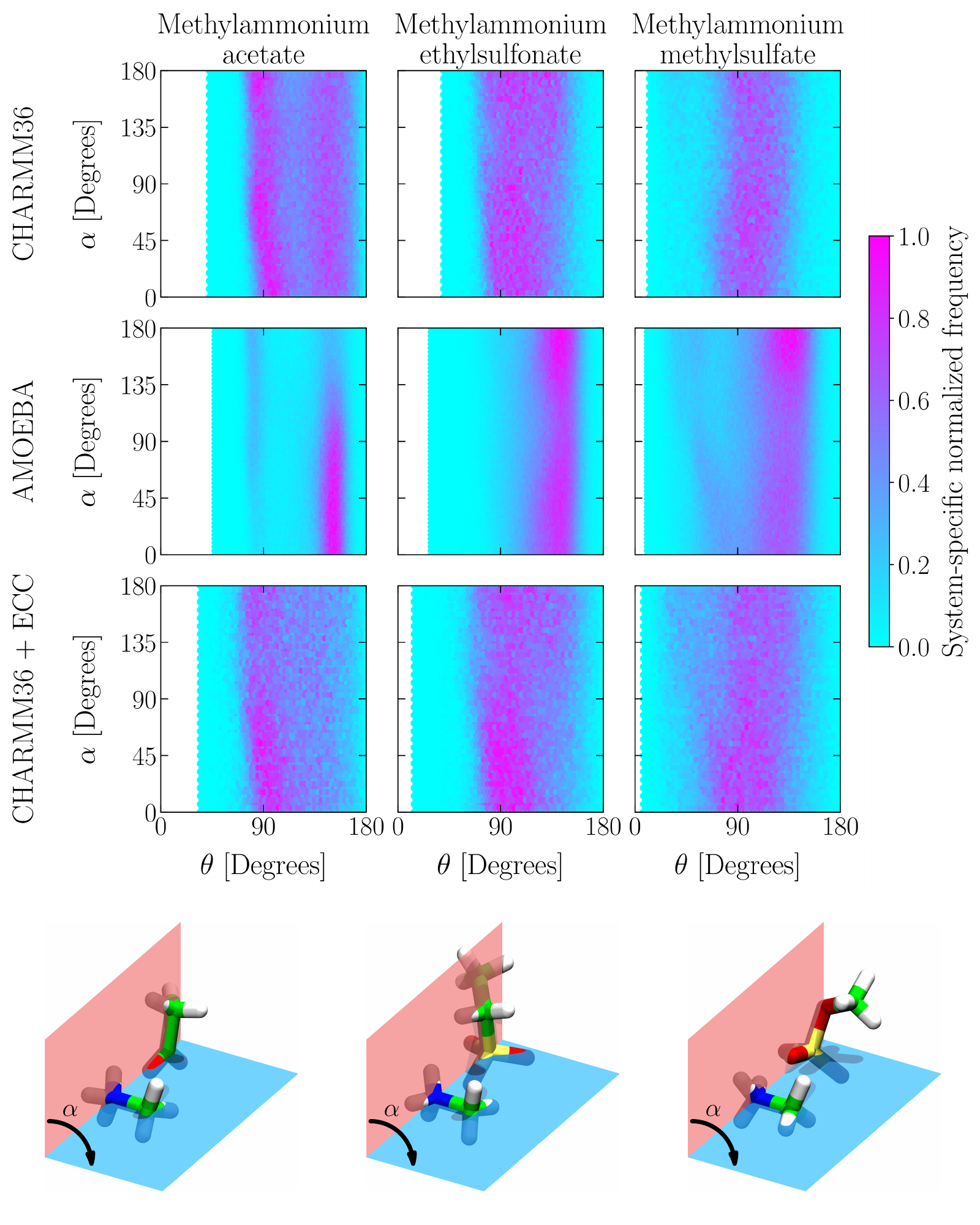}
        \caption{Heat maps of the sampling frequency of anion-methylammonium pairs (columns) with $r < R_{Shell}$ as observed in simulations using different force fields (rows). The normalization of sampling frequency is specific to each system. Refer to Table~\ref{tab:coordinates} for the definitions of $\theta$ and $\alpha$.}
        \label{fig:heatmap_mamm}
    \end{center}
    \end{figure}

     The angular coordinate $\alpha$ was defined to quantify the orientation of cations relative to anions. For the planar guanidinium and imidazolium molecules, values that are closer to 0\textdegree \ indicate anion interactions within the cation plane, whereas values closer to 90\textdegree \ indicate anion interactions orthogonal to the cation plane. For methylammonium, $\alpha$ represents a torsion angle, where values that are closer to 0\textdegree \ indicate the methylammonium $\ce{CH_3}$ group is spatially inline with the anion ``tail" atom that is highlighted in green in Figure~1, whereas values closer to 180\textdegree \ indicate the methylammonium $\ce{CH_3}$ is oriented away from the anion. ZigureS~\ref{fig:heatmap_guan}-\ref{fig:heatmap_mamm} show heat maps in the ($\theta$, $\alpha$) space for associated ion pairs ($r < R_{Shell}$). Figure~\ref{fig:heatmap_guan} shows that, as was observed from Figure~\ref{fig:3D_cations}, anion interactions within the guanidinium plane are generally preferred and this corresponds to the configurational space that in principle should be most accessible during the adsorption of a full protein onto a CEX surface. However for acetate, AMOEBA also has out-of-guanidinium-plane sampling at $\theta \sim 90$\textdegree, which is mimicked imperfectly by the ECC and seems consistent with the findings in Mason et al.\cite{Mason2019a} For imidazolium, in-plane interactions are also preferred but are less distinct (Figure~\ref{fig:heatmap_imim}). For the methylammonium systems there is generally no torsional preference in the CHARMM36 or ECC data, but there are distinct preferences in AMOEBA (Figure~\ref{fig:heatmap_mamm}). For instance the association of acetate with methylammonium favors $\alpha < 90$\textdegree \ such that the $\ce{CH_3}$ groups of both methylammonium and acetate may interact. Conversely there is a slight preference for the methylammonium $\ce{CH_3}$ group to be oriented away from the the sulfur-containing ligands, which represents the configurational space that in principle should be most accessible during protein interactions.
     
    \section{Hydration free energy ($\mu^{\rm{ex}}$) simulations}
    \subsection{Molecular quasichemical theory (mQCT) and methods}

    The hydration free energy of a solute is formally defined by the solute's excess chemical potential $\mu^{\rm{ex}}$, which represents the contribution of the solute's intermolecular interactions in solution to the system Gibbs free energy. According to the potential distribution theorem \cite{Beck2006}
    \begin{eqnarray}
    \beta \mu^{\rm{ex}} = \ln \int e^{\beta \varepsilon} P(\varepsilon) d\varepsilon = \ln \langle e^{\beta \varepsilon} \rangle
    \label{eq:pdt}
    \end{eqnarray}
    where $\beta = 1/k_{\rm B}T$, $k_{\rm B}$ is Boltzmann's constant, $T$ is temperature, $\varepsilon$ is the solute-solvent interaction energy and $P(\varepsilon)$ is the probability density distribution of $\varepsilon$. To make $\mu^{\rm{ex}}$ amenable to estimation from molecular simulation, Equation \ref{eq:pdt} may be partitioned according to molecular quasichemical theory as \cite{Asthagiri2021b, Weber2012, Tomar2016, Adhikari2022, Weber2011}
    \begin{eqnarray}
    \beta \mu^{\rm{ex}} = \underbrace{- \ln p_0[\phi(r; \lambda_G)]}_\text{Packing} + \underbrace{\beta\mu^{\rm{ex}}_{\rm{LR}}[P(\varepsilon \; | \; \phi(r; \lambda_G))]}_\text{Long-range} + \underbrace{\ln x_0[\phi(r; \lambda_G)]}_\text{Chemistry}
    \label{eq:qct}
    \end{eqnarray}
    where $- \ln p_0$ represents packing (primitive hydrophobic) contributions to $\mu^{\rm{ex}}$, which corresponds to the free energy required to open a cavity in the liquid that can accommodate the solute. Long-range solute-solvent interaction energy contributions are represented by $\mu^{\rm{ex}}_{\rm{LR}}$ and $\ln x_0$ refers to the contribution of short-range chemical interactions that occur in the solute's inner hydration shell. Each of these terms are functionals of an imposed potential field $\phi$ that is used to move the solvent interface a distance $\lambda$ away from the solute but the sum of these terms is in principle independent of $\phi$. The probability distribution $P(\varepsilon|\phi)$ is that of $\varepsilon$ when the solvent is conditioned by $\phi$. In this work the soft repulsive WCA potential is used to define $\phi$ as \cite{Tomar2016, Adhikari2022, Weber2012}
    \begin{eqnarray}
    \phi (r_{ij} ; \lambda) = 
    \begin{cases}
        4a \left( \left( \dfrac{b}{r_{ij} - \lambda +  \sqrt[6]{2}b} \right)^{12} - \left( \dfrac{b}{r_{ij} - \lambda +  \sqrt[6]{2}b} \right)^{6} \right) + a, &  r_{ij} < \lambda \\
        0, & r_{ij} \ge \lambda
    \end{cases}
    \label{eq:wca_potential}
    \end{eqnarray}
    with parameters $a = 0.155$~kcal/mol and $b = 3.1655$~\AA\ selected from the SPC/E water model.\cite{Berendsen1987, Chatterjee2008} Here $r_{ij}$ represents the distance between solvent atom $i$ and solute atom $j$; in practice the potential is applied only to the water oxygen atoms based on the position of the solute heavy atoms. The union of spherical shells of radius $\lambda$ that are centered on the solute heavy atoms therefore defines the solute envelope and $\lambda$ may be understood as the range of $\phi$. The upper range is chosen to evacuate the inner hydration shell around the solute of interest such that $P(\varepsilon|\phi)$ may be described using a Gaussian distribution; this upper range is denoted $\lambda_G$ and was taken to be 5 \AA\ based on previous reports. \cite{Asthagiri2017, Adhikari2022}

    The magnitude of the WCA repulsive force is given by
    \begin{eqnarray}
    F(r_{ij} ; \lambda) = -\frac{\partial \phi}{\partial r_{ij}} = 
    \begin{cases}
         \dfrac{-24 a}{b} \left( \dfrac{b}{r_{ij} - \lambda +  \sqrt[6]{2}b} \right)^7 \left(1-2 \left( \dfrac{b}{r_{ij} - \lambda +  \sqrt[6]{2}b} \right)^6 \right), &  r_{ij} < \lambda \\
        0, & r_{ij} \ge \lambda
    \end{cases}
    \label{eq:wca_force}
    \end{eqnarray}
    and the total force exerted by the solvent on the solute-solvent interface at $\lambda$ may be obtained in practice as 
    \begin{eqnarray}
    F_{\rm{Wall}}(\lambda) = \sum_{i=1}^{N_{\rm{Solvent}}} \sum_{j=1}^{N_{\rm{Centers}}} F(r_{ij} ; \lambda)
    \label{eq:fwall}
    \end{eqnarray}
    where the summations in $i$ and $j$ proceed over the number of solvent molecules and the number of heavy atom centers in the solute, respectively. Packing contributions to $\mu^{\rm{ex}}$ are computed as 
    \begin{eqnarray}
    - \ln p_0[\phi(r; \lambda_G)] = \beta \int_{0}^{\lambda_G} \langle F_{\rm{Wall}} (\lambda) \rangle_{0} \; d \lambda
    \label{eq:packing}
    \end{eqnarray}   
    where $\langle F_{\rm{Wall}} (\lambda) \rangle_{0}$ is evaluated in the absence of the solute (i.e., with an uncoupled system) but the solvent is nonetheless conditioned by $\phi$ using fixed representative positions of the solute heavy atoms. Chemistry contributions may be similarly obtained as 
    \begin{eqnarray}
    \ln x_0[\phi(r; \lambda_G)] = -\beta \int_{0}^{\lambda_G} \langle F_{\rm{Wall}} (\lambda) \rangle \; d \lambda
    \label{eq:chemistry}
    \end{eqnarray}   
    where $\langle F_{\rm{Wall}} (\lambda) \rangle$ is evaluated in the presence of the solute (i.e. with a coupled system). In practice this integral may be evaluated over a smaller domain
    \begin{align}
    \ln x_0[\phi(r; \lambda_G)] &= -\beta \left( \int_{0}^{2 \; \text{\AA}} \langle F_{\rm{Wall}} (\lambda) \rangle \; d \lambda + \int_{2 \; \text{\AA}}^{\lambda_G} \langle F_{\rm{Wall}} (\lambda) \rangle \; d \lambda \right) \\
    &\approx -\beta \int_{2 \; \text{\AA}}^{\lambda_G} \langle F_{\rm{Wall}} (\lambda) \rangle \; d \lambda
    \label{eq:chemistry_2A}
    \end{align}
    because it is known from past experience that solvent never enters the domain of $\lambda~<~2.5~\text{\AA}$ when the solute is present \cite{Tomar2016}; integral contributions up to 2~\AA\ are therefore recognized as negligible. Assuming that $\lambda_G$ is sufficiently large to describe $P(\varepsilon|\phi)$ as a Gaussian, the long-range contributions may be obtained from a linear response model as \cite{Tomar2016, Adhikari2022}
    \begin{eqnarray}
    \beta \mu^{\rm{ex}}_{\rm{LR}}[P(\varepsilon \; | \; \phi(r; \lambda_G))] = \beta \langle \varepsilon \rangle_{\phi(\lambda_G)} + \beta^2 \frac{\sigma_{\phi(\lambda_G)}^2}{2}
    \label{eq:long-range}
    \end{eqnarray}
    where $\langle \varepsilon \rangle_{\phi(\lambda_G)}$ and $\sigma_{\phi(\lambda_G)}^2$ represent the mean and variance  of $P(\varepsilon|\phi)$, respectively, which is evaluated in the presence of the solute (i.e., with a coupled system) with $\phi(\lambda_G)$. 

    Equations \ref{eq:packing} and \ref{eq:chemistry_2A} were evaluated using 7-point Gauss–Legendre quadrature \cite{Hummer1996a} over each unit \AA\ such that 35 and 21 points were used in total, respectively, given that $\lambda_G~=~5~\text{\AA}$. Simulations were performed in sequence of increasing $\lambda$ with one ionic solute held in fixed position at the center of a box containing 1500 water molecules (box length $\sim$ 35.6 \AA), and the last configuration of one simulation was used as the starting configuration for the next $\lambda$ point. The Tcl-interface to NAMD and the \texttt{CustomExternalForce} class in OpenMM were used to impose $\phi$ in classical and semiclassical simulations, respectively. Classical simulations of 1~ns were performed at each $\lambda$ point with force data output saved every 50~fs and the last 0.5~ns of data were used for estimation of the mean $F_{\rm{Wall}} (\lambda)$. Analogous semiclassical simulations were run for 0.4~ns with data output every 25~fs and the last 0.3~ns were used for analysis. The standard error of the mean force was estimated using the Friedberg-Cameron algorithm\cite{Friedberg1970, Allen1986} and errors were propagated during integration using standard variance addition rules.
    
    For the evaluation of Equation \ref{eq:long-range} both classical and semiclassical simulations were performed for 1~ns with coordinates saved every 100~fs and the last 0.9~ns of data were used for analysis. As in the \emph{in vacuo} energy analyses, the \texttt{pairInteraction} module in NAMD was used to extract directly the net solute-solvent interaction energy from classical simulation frames. Separate \texttt{.arc} files were generated from semiclassical simulation trajectories for the solute + solvent system as well as the collection of solvent molecules. These were analyzed using the Tinker potential energy program \texttt{analyze} to obtain the total intermolecular interaction energy of the solute + solvent system $U_{N+1}$ as well as the interaction energy among all solvent molecules $U_{N}$, and the solute-solvent interaction energy was obtained as $U_{N+1} - U_{N}$.

    Ewald self-interaction corrections were computed following Hummer et al. and added to $\mu^{\rm{ex}}_{\rm{LR}}$.\cite{Hummer1996} CHARMM36 partial charge distributions were used in estimating Ewald corrections for both classical and semiclassical simulations. We note that these corrections apply to long-range interactions that are not sensitive to the details of the charge distribution (values are given in Table~\ref{tab:mqct_results} and are $-13.18 \pm 0.04$ kcal/mol for all ions in classical simulations) and errors arising from this approximation for semiclassical simulations are expected to be less than the uncertainty in the total $\mu^{\rm{ex}}_{\rm{LR}}$ estimate.

\begin{figure}[H]
    \begin{center}
        \includegraphics[width=0.95\columnwidth]{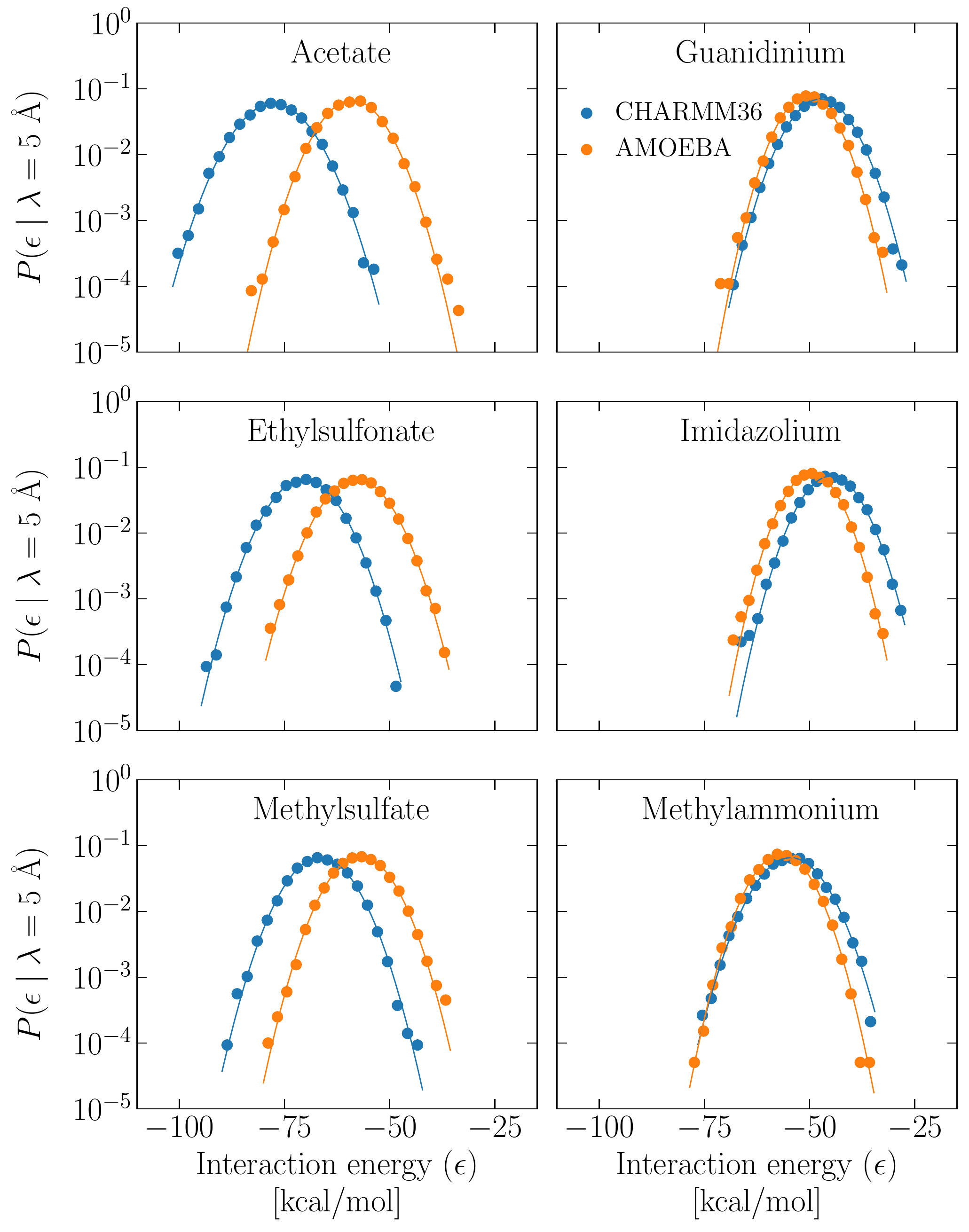}
        \caption{Interaction energy distributions between ions and the solvent when conditioned by $\phi(\lambda_G)$. Points represent the simulation data and lines represent Gaussian fits with the same mean and standard deviation as the data.}
        \label{fig:gaussian_check}
    \end{center}
    \end{figure}

    To validate the applicability of the linear response model (Equation \ref{eq:long-range}), the distribution of interaction energy between the solute and conditioned solvent $P(\varepsilon | \phi(\lambda_G))$ was plotted for each system in Figure~\ref{fig:gaussian_check} alongside its best-fit Gaussian curve. To a good approximation $P(\varepsilon | \phi(\lambda_G))$ is observed to conform to the Gaussian expectation. However, as an orthogonal check, long-range contributions were also computed more rigorously in CHARMM36 using an alchemical transformation to estimate the electrostatic contributions, which are the primary determinant of the final $\mu^{\rm{ex}}_{\rm{LR}}$ value in the absence of polarization. The less substantial van der Waals contributions were still computed using a linear response model, so $\mu^{\rm{ex}}_{\rm{LR}}$ was obtained as
    \begin{equation}
    \beta \mu^{\rm{ex}}_{\rm{LR}}[P(\varepsilon \; | \; \phi(r; \lambda_G))] = \beta \langle \varepsilon \rangle_{\rm{Vdw}, \phi(\lambda_G)} + \beta^2 \frac{\sigma_{\rm{Vdw}, \phi(\lambda_G)}^2}{2} + \beta \int_{0}^{1} \biggl< \sum_{n=1}^{N_{Atoms}} q_n \psi_n (\gamma q_n) \biggr>_{\phi(\lambda_G)} d \gamma
    \label{eq:charging_validation}
    \end{equation}
    where the sum is over all solute atoms, the coupling parameter $\gamma$ scales the atomic partial charges $q_n$ and the electrostatic potential at each solute atom $\psi_n$ is evaluated for scaled partial charges $\gamma q_n$. The values of $\langle \varepsilon \rangle_{\rm{Vdw}, \phi(\lambda_G)}$ and $\sigma_{\rm{Vdw}, \phi(\lambda_G)}$ were estimated using a simulation in which $q_n = 0 \; \forall \; n$. Simulations of 1~ns were performed at 3 Gauss–Legendre quadrature points in $\gamma$ to estimate the integral in Equation \ref{eq:charging_validation}; coordinates were saved every 200~fs and the last 0.8~ns of data were used for analysis. The results are juxtaposed with the linear-response model estimates of $\mu^{\rm{ex}}_{\rm{LR}}$ in Figure~\ref{fig:long_range_method_parity}, showing that the error incurred by using a linear-response model in the reported results is on the order of only a few percent of the $\mu^{\rm{ex}}_{\rm{LR}}$ value.

    \pagebreak
    \begin{landscape}
    \subsection{Supplementary mQCT results}

    \begin{adjustbox}{center, addcode={%
        \begin{minipage}{\width}
        }{
        \caption[Short Caption]{mQCT contributions to ion hydration free energies. All values are in kcal/mol and uncertainties represent the standard error of the mean.}\label{tab:mqct_results}
        \end{minipage}
        }, float=table
        }
    % \begin{adjustbox}{center, caption={mQCT contributions to ion hydration free energies. All values are in kcal/mol and uncertainties represent the standard error of the mean.}, float=table}
    % \label{tab:mqct_results}
    \centering
    \renewcommand{\arraystretch}{0.7}
    \begin{tabular}{c | c | c | c | c | c | c | c } % llllllll
    Force field & Ion & Packing & Uncorrected & Ewald & Corrected & Chemistry & Total \\
     & & & long-range & correction & long-range & & \\ \hline
    CHARMM36    & Acetate        & $22.03 \pm 0.08$ & $-39.86 \pm 0.11$ & $-13.22$ & $-53.08 \pm 0.11$ & $-65.79 \pm 0.08$ & $-96.84 \pm 0.16$ \\
    CHARMM36    & Ethylsulfonate & $27.09 \pm 0.12$ & $-38.68 \pm 0.10$ & $-13.20$ & $-51.87 \pm 0.10$ & $-57.76 \pm 0.08$ & $-82.54 \pm 0.18$ \\
    CHARMM36    & Methylsulfate  & $26.41 \pm 0.11$ & $-35.83 \pm 0.10$ & $-13.16$ & $-48.99 \pm 0.10$ & $-53.37 \pm 0.10$ & $-75.96 \pm 0.18$ \\
    CHARMM36    & Guanidinium    & $21.91 \pm 0.09$ & $-19.99 \pm 0.09$ & $-13.14$ & $-33.13 \pm 0.09$ & $-47.89 \pm 0.08$ & $-59.11 \pm 0.15$ \\
    CHARMM36    & Imidazolium    & $22.32 \pm 0.08$ & $-19.90 \pm 0.09$ & $-13.15$ & $-33.05 \pm 0.09$ & $-37.23 \pm 0.07$ & $-47.96 \pm 0.14$ \\
    CHARMM36    & Methylammonium & $17.57 \pm 0.07$ & $-23.17 \pm 0.09$ & $-13.22$ & $-36.38 \pm 0.09$ & $-39.00 \pm 0.06$ & $-57.82 \pm 0.13$ \\ \hline
    AMOEBA      & Acetate        & $23.92 \pm 0.28$ & $-29.20 \pm 0.19$ & $-13.19$ & $-42.39 \pm 0.19$ & $-67.17 \pm 0.24$ & $-85.64 \pm 0.42$ \\
    AMOEBA      & Ethylsulfonate & $29.76 \pm 0.28$ & $-27.16 \pm 0.21$ & $-13.16$ & $-40.31 \pm 0.21$ & $-58.00 \pm 0.29$ & $-68.55 \pm 0.45$ \\
    AMOEBA      & Methylsulfate  & $29.59 \pm 0.30$ & $-28.60 \pm 0.20$ & $-13.14$ & $-41.74 \pm 0.20$ & $-54.46 \pm 0.23$ & $-66.61 \pm 0.43$ \\
    AMOEBA      & Guanidinium    & $23.29 \pm 0.33$ & $-28.81 \pm 0.13$ & $-13.11$ & $-41.92 \pm 0.13$ & $-43.13 \pm 0.22$ & $-61.76 \pm 0.41$ \\
    AMOEBA      & Imidazolium    & $24.33 \pm 0.25$ & $-28.77 \pm 0.14$ & $-13.12$ & $-41.90 \pm 0.14$ & $-39.18 \pm 0.22$ & $-56.75 \pm 0.36$ \\
    AMOEBA      & Methylammonium & $18.51 \pm 0.29$ & $-32.23 \pm 0.15$ & $-13.19$ & $-45.42 \pm 0.15$ & $-45.31 \pm 0.28$ & $-72.22 \pm 0.43$
    \end{tabular}
    \end{adjustbox}
    \end{landscape}

    \begin{figure}[H]
    \begin{center}
        \includegraphics[width=0.49\columnwidth]{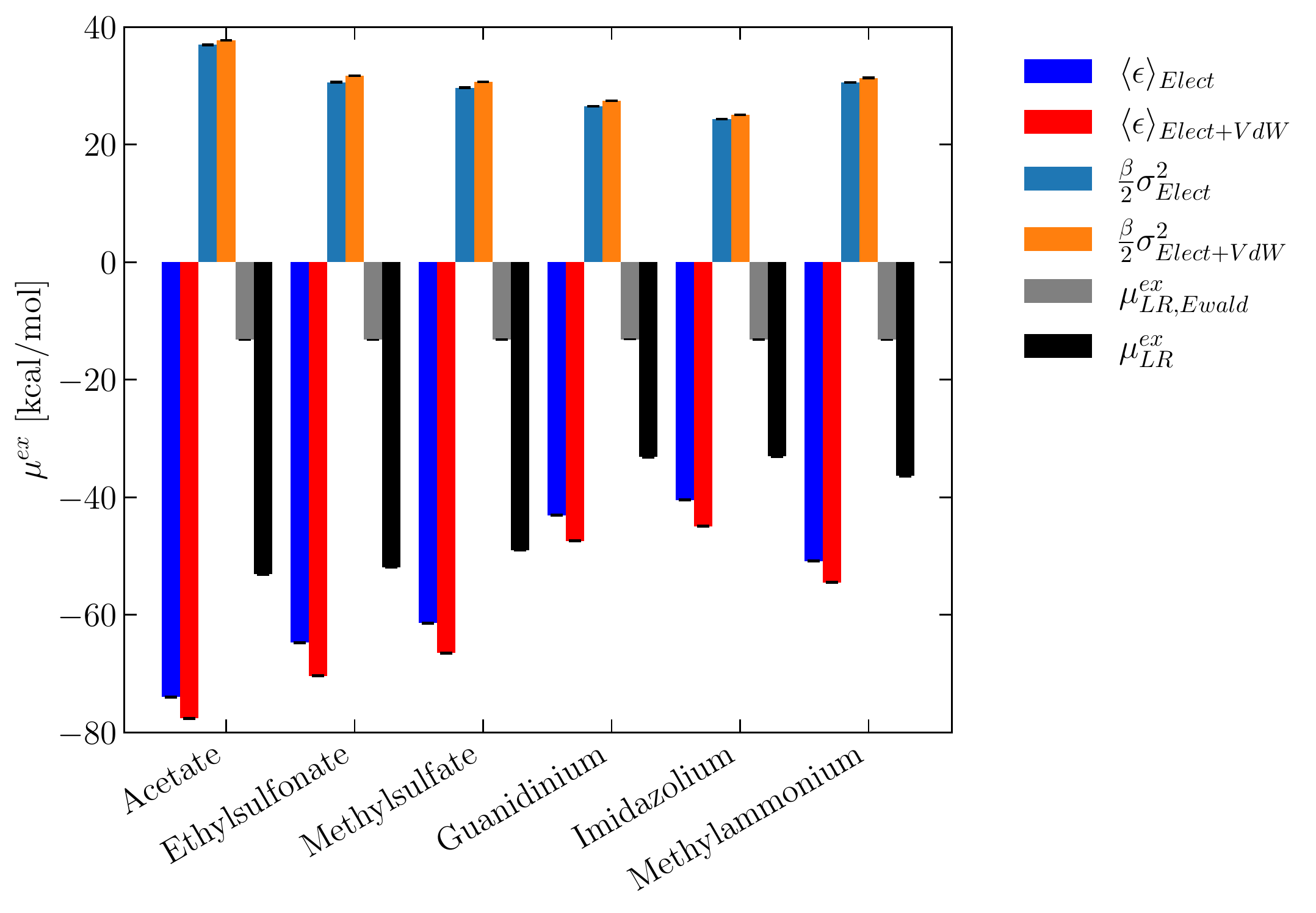}
        \includegraphics[width=0.49\columnwidth]{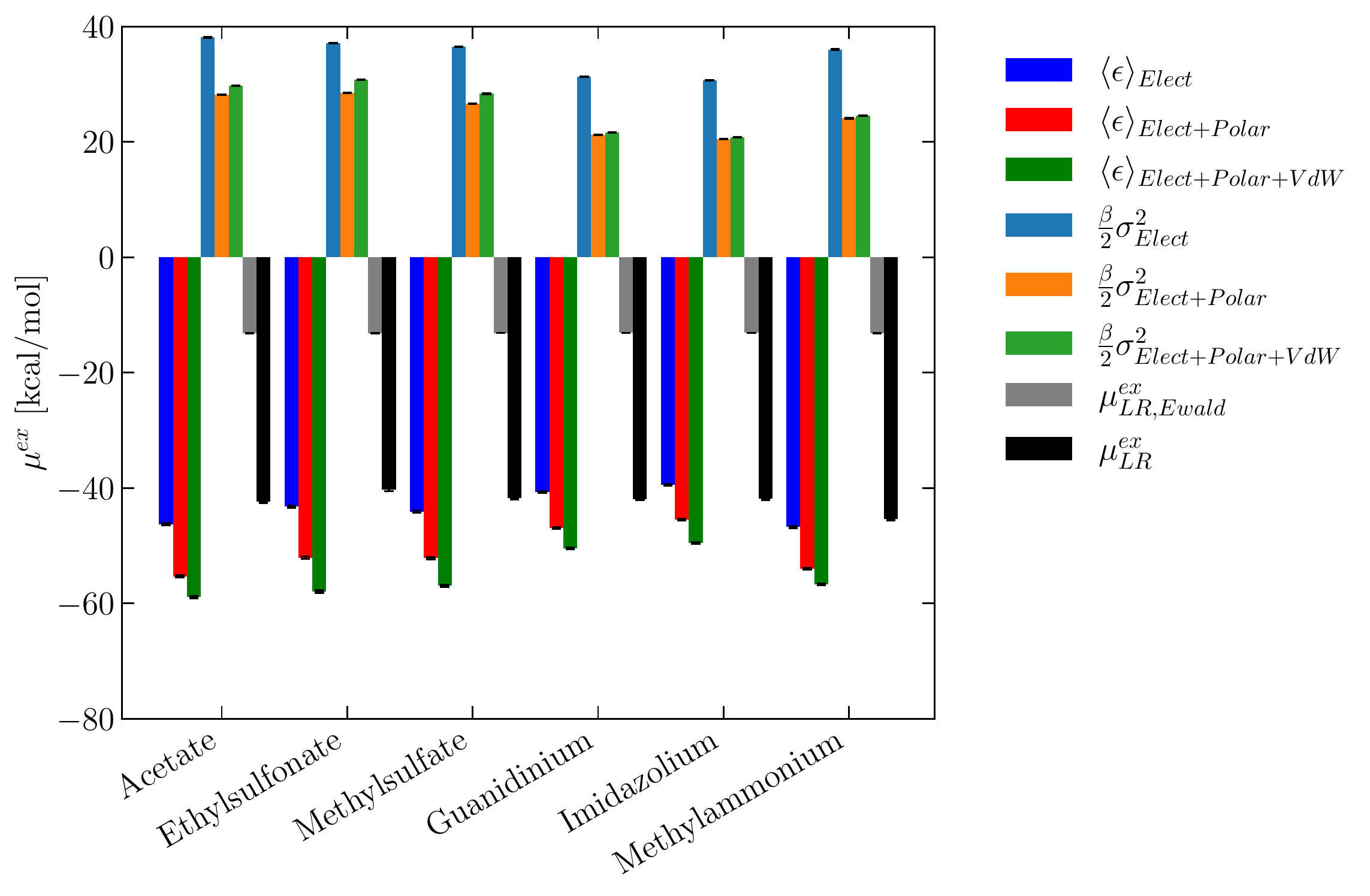}
        \caption{Comparison of the components of long-range contributions to the hydration free energy for CHARMM36 and AMOEBA. Error bars represent $2\times$ the standard error of the mean. The results show that multipole electrostatics itself reduces the difference between long-range components of anions and cations. }
        \label{fig:long_range_compare}
    \end{center}
    \end{figure}
    
    \begin{figure}[H]
    \begin{center}
        \includegraphics[width=0.7\columnwidth]{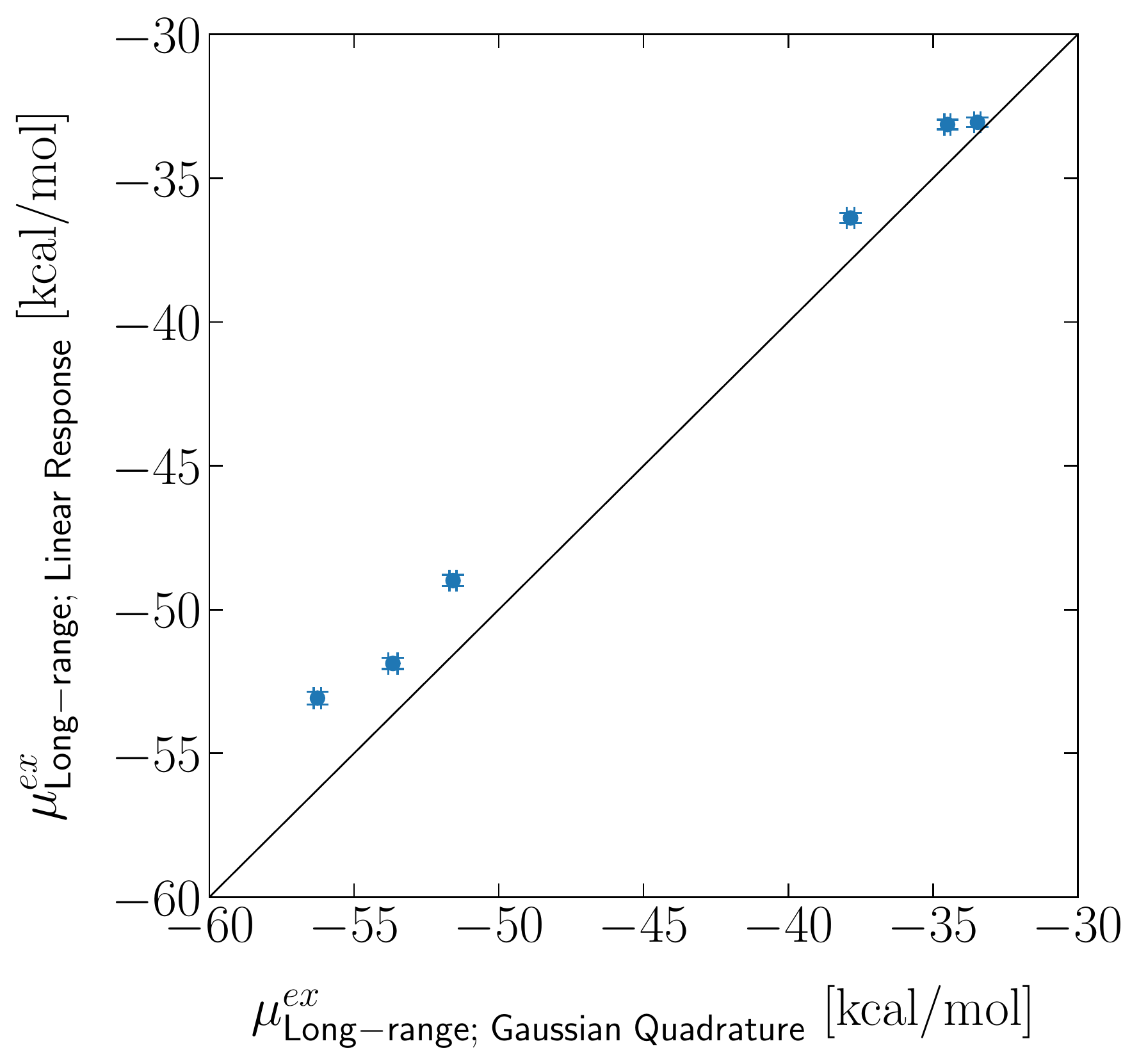}
        \caption{Comparison of Ewald-corrected long-range contributions to the hydration free energy as estimated in CHARMM36 using either 3-point Gaussian quadrature for electrostatic contributions (abscissa, Equation \ref{eq:charging_validation}) or the linear response model (ordinate, Equation \ref{eq:long-range}). Error bars represent $2\times$ the standard error of the mean.}
        \label{fig:long_range_method_parity}
    \end{center}
    \end{figure}

    \clearpage
    \section{Estimation of $\mu^{\rm{ex}}$ from experimental data}
    
    \subsection{Thermodynamic cycle}
    
    \begin{figure}[H]
    \begin{center}
        \includegraphics[width=1\columnwidth]{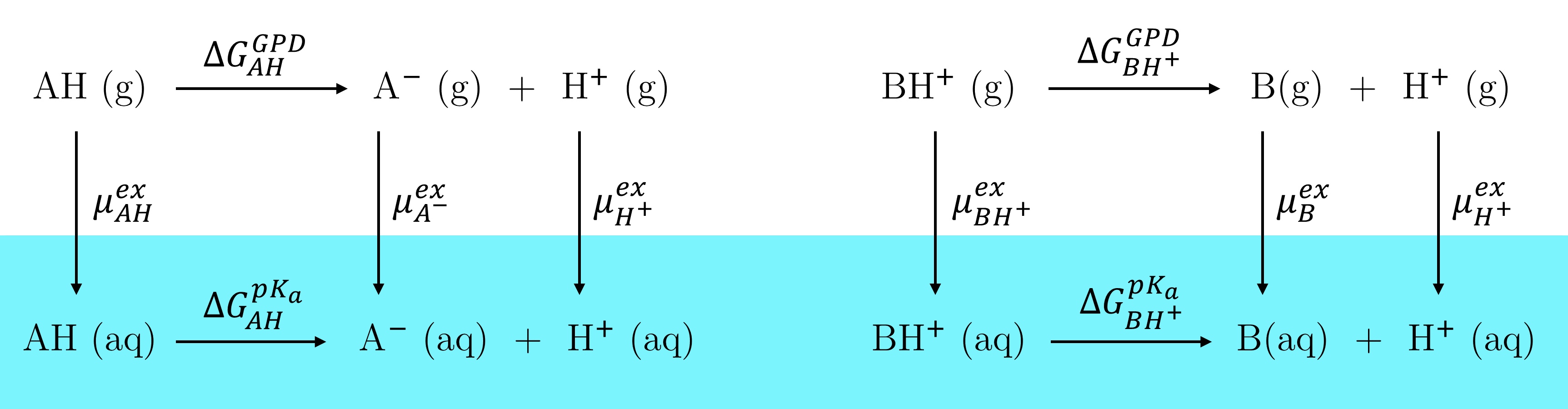}
        \caption{Thermodynamic cycles that were used to estimate anion ($\ce{A^-}$, left) and cation ($\ce{BH^+}$, right) hydration free energies from experimental data.}
        \label{fig:thermo_cycle}
    \end{center}
    \end{figure}
    
    We have used thermodynamic cycles based on proton dissociation (TCPD, Figure~\ref{fig:thermo_cycle})\cite{Fossat2021, Lim1991, Pearson1986} to estimate $\mu^{\rm{ex}}$ from experimental data for comparison with simulation results. Specifically, $\mu^{\rm{ex}}$ was estimated for anions ($\ce{A^-}$) and cations ($\ce{BH^+}$) as
    \begin{equation}
    \mu^{\rm{ex}}_{A^-} = \mu^{\rm{ex}}_{AH} - \Delta G^{\rm{GPD}}_{AH} + \Delta G^{pK_a}_{AH} - \mu^{\rm{ex}}_{H^+} 
    \label{eq:tcpd_anion}
    \end{equation}
    \begin{equation}
    \mu^{\rm{ex}}_{BH^+} = \mu^{\rm{ex}}_{B} + \Delta G^{\rm{GPD}}_{BH^+} - \Delta G^{pK_a}_{BH^+} + \mu^{\rm{ex}}_{H^+} 
    \label{eq:tcpd_cation}
    \end{equation}
    where $\mu^{\rm{ex}}_{AH \; / \; B}$ represents the hydration free energy of the corresponding neutral conjugate $\ce{AH}$ or $\ce{B}$, $\Delta G^{\rm{GPD}}_{AH \; / \; BH^+}$ represents the free energy change of the gas-phase deprotonation of $\ce{AH}$ or $\ce{BH^+}$ (i.e., the gas-phase basicity of the cation conjugate base $\ce{B}$), $\Delta G^{pK_a}_{AH \; / \; BH^+}~=~2.30~RT~pK_a$ represents the corresponding free energy change of titration in the aqueous phase and $\mu^{\rm{ex}}_{H^+}$ represents the proton hydration free energy. In this work $\mu^{\rm{ex}}_{H^+}$ was taken to be $-260.9~\pm~5.8$ kcal/mol, which represents the mean and standard deviation of 72 independent estimates.\cite{Fossat2021} The appreciable uncertainty that surrounds this value is due to the extrathermodynamic assumptions that must be made to deconvolute measurable salt hydration free energies into anion and cation contributions, which are not directly accessible to experiment.\cite{Grossfield2003} Experimental uncertainties were not reported for much of the required data, so a representative uncertainty of 2.5 kcal/mol was assumed for the sum of the first three terms on the right-hand side of equations \ref{eq:tcpd_anion} and \ref{eq:tcpd_cation}, leading to a total uncertainty of 6.3 kcal/mol in the $\mu^{\rm{ex}}$ estimates for anions and cations. Although this is larger than what is typically reported by independent experimental studies, this uncertainty reflects the inherent ambiguities of estimating single ion hydration free energies and using the TCPD with a constant $\mu^{\rm{ex}}_{H^+}$ enables a consistent comparison with simulation results.

    \subsection{Experimental estimates of $\mu^{\rm{ex}}$}
    
    \begin{adjustbox}{center, 
    caption={Ions for which complete experimental data sets exist to estimate $\mu^{\rm{ex}}$ using a thermodynamic cycle based on proton dissociation (TCPD). \cite{Fossat2021, Lim1991, Pearson1986, Reif2012} TCPD estimates are based on a proton hydration free energy of $-260.9 \pm 5.8$ kcal/mol, which represents the mean and standard deviation of 72 independent estimates.\cite{Fossat2021} Experimental uncertainties were not reported for much of the required data, so a representative uncertainty of 2.5 kcal/mol was assumed for the sum of the first three terms on the right-hand side of equations \ref{eq:tcpd_anion} and \ref{eq:tcpd_cation}, leading to a total uncertainty of 6.3 kcal/mol in the $\mu^{\rm{ex}}$ estimates. $\mu^{\rm{ex}}_{AH \; / \; B}$ represents the hydration free energy of the neutral conjugate and $\Delta G^{\rm{GPD}}_{AH \; / \; BH^+}$ represents the free energy change of gas-phase deprotonation.}, float=table}
    \centering
    \label{tab:thermo_cycle_data}
    \renewcommand{\arraystretch}{0.8}
    \begin{threeparttable}
    \begin{tabular}{c | c | c | c | c | c}
    Ion & Neutral conjugate & $\mu^{\rm{ex}}_{AH \; / \; B}$ & $\Delta G^{\rm{GPD}}_{AH \; / \; BH^+}$ & $pK_a$ & TCPD estimate of \\
    (A$^-$ / BH$^+$) & (AH / B)  & [kcal/mol] & [kcal/mol] & & $\mu^{\rm{ex}}_{A^-}$ / $\mu^{\rm{ex}}_{BH^+}$ [kcal/mol] \\ \hline
    Acetate & Acetic acid & $-6.7$ \cite{Cramer1991} & $341.4 \pm 1.9$ \cite{Taft1987, Cumming1977, Fujio1981} & 4.76 \cite{Settimo2014} & $-80.7 \pm 6.3$\\
    Guanidinium & Guanidine & $-11.5$ \cite{Zhang2017}$^{,}$\tnote{a} & 226.9\cite{Hunter1998} & 13.65\cite{Wolfenden1981} & $-64.1 \pm 6.3$\\
    Imidazolium & Imidazole & $-9.63$ \cite{Rizzo2006} & 217.3 \cite{Hunter1998} & 7.05 \cite{Walba1961} & $-62.8 \pm 6.3$\\
    Methylammonium & Methylamine & $-4.56$ \cite{In2005} & 206.6 \cite{Hunter1998} & 10.6 \cite{Lim1991} & $-73.3 \pm 6.3$
    \end{tabular}
    \begin{tablenotes}
        \linespread{1}\small
        \item[a] This is an approximate experimental estimate based on the hydration free energy of methylguanidine ($-11.2$~kcal/mol)\cite{Wolfenden1981}, which was decreased by 0.3~kcal/mol in an attempt to remove the contribution of the methyl group.\cite{Reif2012} 
    \end{tablenotes}
    \end{threeparttable}
    \end{adjustbox}    

    \clearpage

    \begin{adjustbox}{center, 
    caption={Comparison of mQCT and experimental estimates of ion hydration free energies. All values are in kcal/mol. Uncertainties in mQCT results represent the standard error of the mean, whereas the reported uncertainties in experimental estimates represent the measurement standard deviation. Where available, the first entry in the list of experimental data represents the TCPD estimate and the subsequent entries were taken directly from the literature (without correction for any differences in free energy reference values that are related to the underlying extrathermodynamic assumptions).}, float=table}
    \centering
    \label{tab:mqct_lit_comparison}
    \renewcommand{\arraystretch}{0.8}
    \begin{threeparttable}
        \begin{tabular}{c | c | c | c }
        Ion            & CHARMM36 & AMOEBA & Experimental estimates \\ \hline
        Acetate        & $-96.84 \pm 0.16$ & $-85.64 \pm 0.42$ & $-80.7 \pm 6.3$,\tnote{a} \ \ $-75 \pm 2$ \cite{Pearson1986}, \ $-77$ \cite{Cramer1991}, \ $-79 \pm 3$ \cite{Gilson1988}, \ $-89.1$ \cite{Marcus2013} \\
        Ethylsulfonate & $-82.54 \pm 0.18$ & $-68.55 \pm 0.45$ & - \\
        Methylsulfate  & $-75.96 \pm 0.18$ & $-66.61 \pm 0.43$ & - \\
        Guanidinium    & $-59.11 \pm 0.15$ & $-61.76 \pm 0.41$ & $-64.1 \pm 6.3$,\tnote{a} \ \ $-139 \pm 2$ \cite{Marcus2012}$^{,}$\tnote{b} \\
        Imidazolium    & $-47.96 \pm 0.14$ & $-56.75 \pm 0.36$ & $-62.8 \pm 6.3$\tnote{a} \\
        Methylammonium & $-57.82 \pm 0.13$ & $-72.22 \pm 0.43$ & $-73.3 \pm 6.3$,\tnote{a} \ \ $-68 \pm 2$ \cite{Pearson1986}, \ $-70$ \cite{Cramer1991}, \ $-71\pm 3$ \cite{Gilson1988}    
        \end{tabular}
        \begin{tablenotes}
            \linespread{1}\small
            \item[a] Estimate based on the thermodynamic cycle (Table~S4).
            % Hard coded
            \item[b] Some controversy surrounds this measurement\cite{Houriez2017} and quantum mechanics estimates ($-57.9~\pm~0.8$ kcal/mol)\cite{Gokcen2014} are much closer to the values estimated using CHARMM36, AMOEBA and the thermodynamic cycle.            
        \end{tablenotes}
    \end{threeparttable}
    \end{adjustbox}

    \clearpage
   %  \bibliography{myref}
  
  \providecommand{\latin}[1]{#1}
\makeatletter
\providecommand{\doi}
  {\begingroup\let\do\@makeother\dospecials
  \catcode`\{=1 \catcode`\}=2 \doi@aux}
\providecommand{\doi@aux}[1]{\endgroup\texttt{#1}}
\makeatother
\providecommand*\mcitethebibliography{\thebibliography}
\csname @ifundefined\endcsname{endmcitethebibliography}
  {\let\endmcitethebibliography\endthebibliography}{}